\documentclass[aps,prl,a4paper,superscriptaddress,twocolumn,showpacs,amsmath,amssymb]{revtex4-1}

\usepackage{graphicx,epsfig}
\usepackage[usenames]{color}

\usepackage[english]{babel}

\allowdisplaybreaks
\begin{document}

\newcommand{\cau}{\underline{c}^{\phantom{\dagger}}}
\newcommand{\ccu}{\underline{c}^\dagger}

\newcommand{\hu}{\underline{\hat{H}}}

\newcommand{\Dp}{\hat{\Delta}_{\text{p}}}
\newcommand{\Dh}{\hat{\Delta}_{\text{h}}}
\newcommand{\Deltap}{[\hat{\Delta}_{\text{p}}]}
\newcommand{\Deltah}{[\hat{\Delta}_{\text{h}}]}
\newcommand{\R}{\mathcal{R}}
\newcommand{\Rh}{\hat{\mathcal{R}}}

\newcommand{\uA}{|\underline{A},Ri\rangle}
\newcommand{\uB}{|\underline{B},R'j\rangle}

\newcommand{\E}{\mathcal{E}}
\newcommand{\G}{\mathcal{G}}
\newcommand{\Lag}{\mathcal{L}}
\newcommand{\M}{\mathcal{M}}
\newcommand{\N}{\mathcal{N}}
\newcommand{\U}{\mathcal{U}}
\newcommand{\F}{\mathcal{F}}
\newcommand{\V}{\mathcal{V}}
\newcommand{\C}{\mathcal{C}}
\newcommand{\I}{\mathcal{I}}
\newcommand{\s}{\sigma}
\newcommand{\up}{\uparrow}
\newcommand{\dw}{\downarrow}
\newcommand{\h}{\hat{H}}
\newcommand{\himp}{\hat{H}_{\text{imp}}}
\newcommand{\g}{\mathcal{G}^{-1}_0}
\newcommand{\D}{\mathcal{D}}
\newcommand{\A}{\mathcal{A}}
\newcommand{\projs}{\hat{\mathcal{S}}_d}
\newcommand{\proj}{\hat{\mathcal{P}}_d}
\newcommand{\K}{\textbf{k}}
\newcommand{\Q}{\textbf{q}}
\newcommand{\T}{\tau_{\ast}}
\newcommand{\io}{i\omega_n}
\newcommand{\eps}{\varepsilon}
\newcommand{\+}{\dag}
\newcommand{\su}{\uparrow}
\newcommand{\giu}{\downarrow}
\newcommand{\0}[1]{\textbf{#1}}
\newcommand{\ca}{c^{\phantom{\dagger}}}
\newcommand{\cc}{c^\dagger}
\newcommand{\aaa}{a^{\phantom{\dagger}}}
\newcommand{\aac}{a^\dagger}
\newcommand{\bba}{b^{\phantom{\dagger}}}
\newcommand{\bbc}{b^\dagger}
\newcommand{\da}{d^{\phantom{\dagger}}}
\newcommand{\dc}{d^\dagger}
\newcommand{\fa}{f^{\phantom{\dagger}}}
\newcommand{\fc}{f^\dagger}
\newcommand{\ha}{h^{\phantom{\dagger}}}
\newcommand{\hc}{h^\dagger}
\newcommand{\be}{\begin{equation}}
\newcommand{\ee}{\end{equation}}
\newcommand{\bea}{\begin{eqnarray}}
\newcommand{\eea}{\end{eqnarray}}
\newcommand{\ba}{\begin{eqnarray*}}
\newcommand{\ea}{\end{eqnarray*}}
\newcommand{\dagga}{{\phantom{\dagger}}}
\newcommand{\bR}{\mathbf{R}}
\newcommand{\bQ}{\mathbf{Q}}
\newcommand{\bq}{\mathbf{q}}
\newcommand{\bqp}{\mathbf{q'}}
\newcommand{\bk}{\mathbf{k}}
\newcommand{\bh}{\mathbf{h}}
\newcommand{\bkp}{\mathbf{k'}}
\newcommand{\bp}{\mathbf{p}}
\newcommand{\bL}{\mathbf{L}}
\newcommand{\bRp}{\mathbf{R'}}
\newcommand{\bx}{\mathbf{x}}
\newcommand{\by}{\mathbf{y}}
\newcommand{\bz}{\mathbf{z}}
\newcommand{\br}{\mathbf{r}}
\newcommand{\Ima}{{\Im m}}
\newcommand{\Rea}{{\Re e}}
\newcommand{\Pj}[2]{|#1\rangle\langle #2|}
\newcommand{\ket}[1]{\vert#1\rangle}
\newcommand{\bra}[1]{\langle#1\vert}
\newcommand{\setof}[1]{\left\{#1\right\}}
\newcommand{\fract}[2]{\frac{\displaystyle #1}{\displaystyle #2}}
\newcommand{\Av}[2]{\langle #1|\,#2\,|#1\rangle}
\newcommand{\av}[1]{\langle #1 \rangle}
\newcommand{\Mel}[3]{\langle #1|#2\,|#3\rangle}
\newcommand{\Avs}[1]{\langle \,#1\,\rangle_0}
\newcommand{\eqn}[1]{(\ref{#1})}
\newcommand{\Tr}{\mathrm{Tr}}

\newcommand{\Vb}{\bar{\mathcal{V}}}
\newcommand{\Vd}{\Delta\mathcal{V}}
\def\P{P_{02}}
\newcommand{\Pb}{\bar{P}_{02}}
\newcommand{\Pd}{\Delta P_{02}}
\def\t{\theta_{02}}
\newcommand{\tb}{\bar{\theta}_{02}}
\newcommand{\td}{\Delta \theta_{02}}
\newcommand{\Rb}{\bar{R}}
\newcommand{\Rd}{\Delta R}

\def\changemargin#1#2{\list{}{\rightmargin#2\leftmargin#1}\item[]}
\let\endchangemargin=\endlist

\title{Slave Boson  Theory of Orbital Differentiation with Crystal Field Effects: \\ Application to UO$_2$}

\author{Nicola Lanat\`a}
\affiliation{Department of Physics and National High Magnetic Field Laboratory, Florida State University, Tallahassee, Florida 32306, USA}
\author{Yongxin Yao}
\affiliation{Ames Laboratory-U.S. DOE and Department of Physics and Astronomy, Iowa State 
	University, Ames, Iowa IA 50011, USA}
\author{Xiaoyu Deng}
\affiliation{Department of Physics and Astronomy, Rutgers University, Piscataway, New Jersey 08856-8019, USA}
\author{Vladimir Dobrosavljevi\'c}
\affiliation{Department of Physics and National High Magnetic Field Laboratory, Florida State University, Tallahassee, Florida 32306, USA}
\author{Gabriel Kotliar}
\affiliation{Department of Physics and Astronomy, Rutgers University, Piscataway, New Jersey 08856-8019, USA}
\affiliation{Condensed Matter Physics and Materials Science Department, Brookhaven National Laboratories, Upton, NY 11973-5000, USA}
\date{\today} 

\begin{abstract}
	
	We derive an exact operatorial reformulation of the rotational invariant slave boson method
	and we apply it to describe the orbital differentiation in strongly correlated electron
	systems starting from first principles.  
	The approach enables us to treat strong electron correlations, spin-orbit coupling and crystal field splittings on the same footing by exploiting the gauge invariance of the mean-field equations. 
	We apply our theory to the archetypical nuclear fuel UO$_2$, and show that the ground state of this system displays a pronounced orbital differention within the $5f$ manifold, with 
	Mott localized $\Gamma_8$ and extended $\Gamma_7$ electrons.
	
\end{abstract}

\maketitle

Orbital differentiation, where states with different orbital character exhibit different
levels of correlation, is a pervasive phenomena in condensed matter systems~\cite{orbital-selective-Mott-1,orbital-selective-Mott-2,XiDai_Mott-IntegerOccupation,FeSe-FeTe}, 
which
gives rise to multiple functionalities in strongly correlated multiorbital systems.  
In all known Mott systems in nature only a fraction of electrons form localized
magnetic moments, while the other electronic states 
are extended (but away from the Fermi level).
These systems are commonly called ``selective Mott insulators'', and the transition into 
these states is called ``orbitally selective Mott transition''.
Understanding the mechanism driving the selection process is a fundamental
question in condensed matter.
This issue is especially nontrivial to address in low-symmetry $5f$ electron systems,
where the competition between inter- and intra-orbital interactions, the crystal field splittings (CFS)
and the spin-orbit coupling (SOC) is very complicated, as none of these energy scales is negligible.
Orbital differentiation is also a key issue in the presence of disorder~\cite{Vlad1,Marianetti-disorder}
and/or charge ordering (Wigner-Mott transitions~\cite{Vlad5}),
where only a fraction of the electrons Mott-localize.
Addressing these issues quantitatively and in an unbiased ``ab-initio''
fashion is very challenging.
In this work we address the orbital differentiation problem from an ab-initio
perspective
%
%
%
using the rotationally invariant 
slave boson (RISB) mean-field
theory~\cite{Kotliar-Ruckenstein, Georges,rotationally-invariant_SB}. 
As we demonstrate, this method can be derived from 
an \emph{exact} operatorial reformulation of the many-body problem,
which 
reproduces
the Gutzwiller approximation~\cite{Gutzwiller3} at the mean-field
level~\cite{equivalence_GA-SB,lanata} and constitutes 
a starting point to calculate further corrections. 
By exploiting the gauge symmetry of the RISB theory, we 
build efficient systematic algorithms which enable us
to solve the mean-field equations and
elucidate the pattern of orbital differentiation 
even in low-symmetry $5f$ electron systems.
We apply this method to UO$_2$~\cite{UO2-Gabi} (the most widely used nuclear fuel),
and provide new insight into the role of the CFS
in the orbital differentiation and 
the nature of the chemical bonds in this material.

\emph{The multi-band Hubbard model:---}
Let us consider a generic multi-band Hubbard model:
\be
\vspace{-0.035cm}
\h\!=\!\sum_{k}\!\sum_{ij=1,..,n_a}
\,\sum_{\alpha=1,..,M_i}
\sum_{\beta=1,..,M_j}
\!\!\!\!
\epsilon^{\alpha\beta}_{k,ij}\, \cc_{ki\alpha}\ca_{kj\beta}
\!+\!\h^{\text{loc}}\!,
\label{hamgen}
\vspace{-0.035cm}
\ee
where 
$k$ is the momentum conjugate to the unit-cell label $R$,
the $n_a$ atoms within the unit cell are labeled by $i,j$,
and the spin-orbitals are labeled by $\alpha,\beta$.
As in Refs.~\onlinecite{Georges,Our-PRX},
the local interaction and the on-site energies are both
included within the definition of:
\be
\vspace{-0.035cm}
\h^{\text{loc}} \equiv \sum_{Ri}\sum_{AB}
\big[H^{\text{loc}}_i\big]_{AB}\,\ket{A,Ri}
\bra{B,Ri}\,,
\label{hloc_fermions}
\vspace{-0.035cm}
\ee
where $|A,Ri\rangle$ are local Fock states:
\be
\vspace{-0.035cm}
|A,Ri\rangle = 
[c^\dagger_{Ri 1}]^{\nu_{1}(A)}\!\!\!\!\!\!.\,.\,.\;
[c^\dagger_{Ri M_i}]^{\nu_{M_i}(A)}\,|0\rangle
\,,
\label{gammastates-app}
\vspace{-0.035cm}
\ee
and $A=1,..,2^{M_i}$ runs over all of the possible lists
of occupation numbers
$\{\nu_1(A),..,\nu_{M_i}(A)\}$.
In particular, in this work we have used
the Slater-Condon parametrization of the on-site
interaction~\cite{LDA+U}.

\emph{Slave Boson reformulation:---}
Here we derive the RISB gauge theory and show that it
constitutes an exact reformulation of
the generic Hubbard system defined above.
As in Ref.~\onlinecite{Georges}, we introduce a new set of fermionic modes
$\{\fa_{Ria}|a=1,..,M_i\}$, that we call quasi-particle operators.
Furthermore, we introduce a bosonic mode $\Phi^\dagga_{RiAn}$
for each couple of fermionic local multiplets
$(|A,Ri\rangle,|n,Ri\rangle)$ having equal number of electrons, i.e., 
$N_A\equiv \sum_{a=1}^{M_i}\nu_{a}(A) = N_n\equiv \sum_{a=1}^{M_i}\nu_{a}(n)$.
Applying the algebra generated by $\{\Phi^\dagger_{RiAn}\}$ and $\{f^\dagger_{Ri a}\}$
to the vacuum $\ket{0}$ generates a new
Fock space $\mathcal{H}_{\text{SB}}$.
We define ``physical Hilbert space''
the subspace $h_{\text{SB}}$ of $\mathcal{H}_{\text{SB}}$
satisfying the following equations 
(Gutzwiller constraints):
\bea
\vspace{-0.035cm}
K^0_{Ri}\!&\equiv&\! \sum_{An} \Phi^\dagger_{Ri An}\!\Phi^\dagga_{Ri An} - \hat{I} \!=\! 0
\label{C1}
\\[-0.5mm]
K_{Riab}\!&\equiv&\!
\label{C2}
\fc_{Ria}\fa_{Rib} \!-\!
\sum_{Anm} [F^\dagger_{ia}F^\dagga_{ib}]_{mn}\,
\Phi^\dagger_{Ri An}\!\Phi^\dagga_{Ri Am}\!=\!0\,,~~~~
\vspace{-0.03cm}
\eea
where
$\hat{I}$ is the identity,
$\left[F_{ia}\right]_{nm}\equiv\langle n,Ri|\,\fa_{Ria}\,| m,Ri\rangle$,
and $\ket{n,Ri}$ and $\ket{m,Ri}$ are Fock states constructed
as in Eq.~\eqref{gammastates-app},
but using the quasi-particle operators $\fa_{Ria}$.

In Ref.~\onlinecite{Our-PRX} it was shown that
the following Hamiltonian is an exact representation of $\h$
within $h_{\text{SB}}$:
\be
\vspace{-0.035cm}
\hu=\!\!\!
\sum_{kij\alpha\beta}\!\!\!\epsilon^{\alpha\beta}_{k,ij}\, \ccu_{ki\alpha}\cau_{kj\beta}
+\!\!\sum_{RiAB}
\!\![H_i^{\text{loc}}]_{AB}\!\sum_n\Phi^\dagger_{RiAn}\!\Phi^\dagga_{RiBn}
,
\label{hu}
\vspace{-0.035cm}
\ee
where $\ccu_{Ri\alpha}\equiv\sum_a \Rh_{Ria\alpha}\,\fc_{Ri a}$,
and the operators
\bea
\vspace{-0.035cm}
\Rh_{Ri a\alpha}=
\sum_{AB}\sum_{nm}
\frac{[F^\dagger_{i \alpha}]_{AB}[F^\dagger_{i a}]_{nm}}{\sqrt{N_A(M_i-N_B)}}\,
\Phi^\dagger_{Ri An} \Phi^\dagga_{Ri Bm}
\label{old_R}
\vspace{-0.035cm}
\eea
are such that $\ccu_{Ri\alpha}$ are a representation in
$h_{\text{SB}}$ of $\cc_{Ri\alpha}$.
A remarkable property of $\hu$ is that 
it is invariant with respect to the gauge Lie group
generated by the Gutzwiller constraint operators $K_{Riab}$,
see Eq.~\eqref{C2}:
\be
\vspace{-0.035cm}
e^{i\sum_{Riab}\theta_{ab}K_{Riab}}
\, \hu\, 
e^{-i\sum_{Riab}\theta_{ab}K_{Riab}}
=\hu\;\,\forall\,\theta=\theta^\dagger
\label{operatorial-gauge-invariance}
\,.
\vspace{-0.035cm}
\ee
In fact, Eq.~\eqref{operatorial-gauge-invariance}
does not hold only within the subspace $h_{\text{SB}}$
(which would be a trivial consequence of Eq.~\eqref{C2}),
but in the entire RISB Fock space
$\mathcal{H}_{\text{SB}}$~\cite{supplemental_material}.

\emph{Operatorial formulation of RISB theory:---}
The operators $\Rh_{Ria\alpha}$ defined above are constructed 
in such a way that $\ccu_{Ri\alpha}$ are a representation in the
physical RISB subspace of the corresponding
original fermionic operators $\cc_{Ri\alpha}$.
However, this construction is \emph{not} unique.
In particular, Eq.~\eqref{old_R} can be modified
as follows:
\be
\vspace{-0.035cm}
\Rh_{Ria\alpha} \!\!=
:\!\sum_{AB}\!\sum_{nm}\!\!
\frac{[F^\dagger_{i \alpha}]_{AB}[F^\dagger_{i a}]_{nm}}{\sqrt{N_A(M_i-N_B)}}\,
\Phi^\dagger_{Ri An}
[\hat{1}\!+\!\hat{X}_{AB}]\,
\Phi^\dagga_{Ri Bm}\!:
\label{new_R}
\vspace{-0.035cm}
\ee
where ``$:$'' indicates the normal ordering~\cite{SB-normal-ordering}, 
and $\hat{X}_{AB}$ is any
normally-ordered algebraic
combination of bosonic ladder operators such that each term contains
at least 2 modes.
In fact, since $\hat{X}_{AB}$ is normally-ordered and
the physical RISB states contain only one boson by construction,
see Eq.~\eqref{C1},
the matrix elements of Eqs.~\eqref{old_R} and \eqref{new_R} are
independent of $\hat{X}_{AB}$ within $h_{\text{SB}}$.

Of course, any choice of $\hat{X}_{AB}$ in Eq.~\eqref{new_R}
would be equivalent if we were able to solve $\hu$ exactly.
However, this choice affects the RISB mean-field approximation
(that we are going to introduce below).
Interestingly, it is possible to construct $\hat{X}_{AB}$
in such a way that: (i) the RISB mean-field theory is exact for
any uncorrelated Hubbard Hamiltonian, and (ii) 
the invariance property [Eq.~\eqref{operatorial-gauge-invariance}]
of $\hu$ with respect to the gauge group remains valid.
To the best of our knowledge, this operatorial construction, which
is derived in the supplemental material
of this work~\cite{supplemental_material}, was not provided in any previous work.

\emph{RISB mean-field theory:---}
At zero temperature, the RISB mean-field theory consists in minimizing
the expectation value of $\hu$ with respect to
$\ket{\Psi_{\text{MF}}} = \ket{\Psi_0}\otimes\ket{\phi}$,
where $\ket{\Psi_0}$ is a Slater determinant constructed with
the quasi-particle operators $\fa_{Ria}$, $\ket{\phi}$
is a bosonic coherent state, and the Gutzwiller constraints,
see Eqs~\eqref{C1} and \eqref{C2}, are enforced only in average.

It can be verified that taking the expectation value of 
Eqs.~\eqref{C1} and \eqref{C2} with respect to
$\ket{\Psi_{\text{MF}}}$ gives:
\bea
\vspace{-0.035cm}
\Tr\big[\phi^\dagger_{i}\phi^\dagga_{i}\big] \!&=&\! 1 
\;\forall\,i
\label{avC1}
\\[-0.5mm]
\left[\Delta_{pi}\right]_{ab} \!\equiv\!
\Tr\big[\phi^\dagger_{i}\phi^\dagga_{i}F^\dagger_{ia}F^\dagga_{ib}\big]
\!&=&\! \Av{\Psi_0}{\!\fc_{Ria}\fa_{Rib}\!}
\;\forall\,i
\,,
\label{avC2}
\vspace{-0.035cm}
\eea
where the matrix elements $[\phi_i]_{An}$, which we call
``slave boson amplitudes'',
are the eigenvalues of the annihilation
operators $\Phi_{RiAn}$ with respect to 
$\ket{\phi}$.
Similarly, it can be verified that
the expectation value of $\hu$ with respect to $\ket{\Psi_{\text{MF}}}$ 
(normalized to the number of $k$-points $\mathcal{N}$)
is given by:
\bea
\vspace{-0.035cm}
\mathcal{E}&\equiv&\frac{1}{\mathcal{N}}\,
\Av{\Psi_{\text{MF}}}{\underline{\h}}
=\sum_i\Tr\big[\phi_i^\dagga\phi_i^\dagger\, H^{\text{loc}}_i\big]
\nonumber\\[-0.5mm]
&+&
\frac{1}{\mathcal{N}}\sum_{kij}\sum_{ab}
\big[\R^\dagga_i\epsilon_{k,ij}\R^\dagger_j\big]_{ab}
\Av{\Psi_0}{\fc_{kia}\fa_{kjb}}
\,,
\label{variationalenergy}
\vspace{-0.035cm}
\eea
where $[\R_i]_{a\alpha}\equiv \Av{\phi}{\Rh_{Ri a\alpha}}$
is given by:
\bea
\vspace{-0.035cm}
[\R_i]_{a\alpha} =
\Tr\big[\phi_i^\dagger F^\dagger_{i\alpha}\phi^\dagga_i F^\dagga_{ib}\big]
\big[
\Delta_{p i}(1-[\Delta_{p i}])\big]^{-\frac{1}{2}}_{ba}
\,,
\label{def-R-mf}
\vspace{-0.035cm}
\eea
$1$ is the identity matrix,
and $\Rh_{Ri a\alpha}$ are the renormalization operators
represented in Eq.~\eqref{new_R}, and constructed explicitly in the
supplemental material~\cite{supplemental_material}.
%
The RISB mean-field theory amounts
to minimize Eq.~\eqref{variationalenergy}
with respect to $\ket{\Psi_{\text{MF}}}$
while fulfilling Eqs.~\eqref{avC1} and \eqref{avC2}.

\emph{Advantages of the gauge invariant formulation:---}
As shown in the supplemental material~\cite{supplemental_material},
the above constrained minimization problem
can be conveniently cast --- analogously to DMFT~\cite{DMFT,Anisimov_DMFT,CDMFT-Lichtenstein} --- 
as a root problem for the variables
$\left(\R_{i},\lambda_{i}\right)$,
where $\R_{i}$ were defined in Eq.~\eqref{def-R-mf}, and
$\lambda_i$ are matrices of Lagrange multipliers introduced in order to
enforce the Gutzwiller constraints [Eq.~\eqref{avC2}].
These variables encode the so called
``Gutzwiller self energy'' of each inequivalent atom, that is defined as:
\be
\vspace{-0.035cm}
\Sigma_i(\omega)\equiv
(I-\R^\dagger_i\R_i^\dagga)(\R^\dagger_i\R_i^\dagga)^{-1}\,\omega
+ (\R_i^{-1}\lambda_i\R_i^{\dagger\,-1})\,,
\label{se}
\vspace{-0.035cm}
\ee
where $Z_i\equiv\R_i^\dagger\R_i^\dagga$ are \emph{matrices}
of quasi-particle weights.
Let us represent formally 
the above-mentioned 
root problem as follows:
\be
\vspace{-0.035cm}
\mathcal{F}[(\R_{1},\lambda_{1}),...,(\R_{n_a},\lambda_{n_a})]=0
\,,
\label{roothubb}
\vspace{-0.035cm}
\ee
where $n_a$ is the number of inequivalent atoms within the unit cell.
As shown in the supplemental material~\cite{supplemental_material},
each evaluation of $\mathcal{F}$ requires to solve $n_a$ impurity models,
where the bath has the same dimension of the impurity for each
inequivalent atom~\cite{Our-PRX}.
An important advantage of the present formulation
with respect to Ref.~\onlinecite{Our-PRX}
is that, by virtue of Eq.~\eqref{operatorial-gauge-invariance},
Eq.~\eqref{roothubb} has a manifold of
physically equivalent solutions, which are mapped one into the other by
the following group of gauge transformations:
$\R_i \rightarrow u_i^\dagger(\theta_i)\,\R_i$,
$\lambda_i \rightarrow u_i^\dagger(\theta_i)\,\lambda_i\,u_i(\theta_i)$,
where $u_{i}(\theta_{i})\equiv e^{i\theta_{i}}$
are generic unitary matrices.
This property effectively reduces the dimension
of the root problem, which makes
the code more stable and speeds
up the convergence by reducing substantially 
the number of evaluations of $\mathcal{F}$ necessary to
solve Eq.~\eqref{roothubb}.
Remarkably, we found that exploiting the
gauge freedom mentioned above
is essential in order to study 
strongly correlated
materials where the SOC and the CFS are equally important,
which generally makes the structure of $\Sigma_i(\omega)$ particularly
complex~\footnote{The interplay between SOC and CFS
	can generate multiple equivalent representations of the point symmetry group
	in the local single particle space,
	so that $\Sigma(\omega)$ is not made automatically
	diagonal by selection rules~\cite{Wigner}.}.
Further technical details are discussed in the
supplemental material~\cite{supplemental_material}.

\emph{Calculations of UO$_2$:---}
UO$_2$ is  widely used as a nuclear fuel.
At ambient pressure it is a Mott insulator and crystallizes in a cubic fluorite structure. 
Given the importance of this material, its electronic structure and energetics
have been extensively investigated both experimentally
and theoretically, e.g., with DFT+U~\cite{DFT+U_UO2_1,DFT+U_UO2_2,DFT+U_UO2_3}
and other single-particle approaches~\cite{hybrid_UO2_1,hybrid_UO2_2}.
However, within these techniques it is not possible to
address the properties of the paramagnetic state of this material, which is stable
above the N\'eel temperature $T_{\text{N}}\simeq 30.8\,K$~\cite{UO2-Neel}.
Because of this reason,
several DMFT studies of paramagnetic UO$_2$ have been recently performed~\cite{UO2-Gabi,UO2-Werner,UO2-Savrasov,DMFT-UO2-referee}.
A particularly important statement concerning the orbital differentiation of the U-$5f$ electrons 
was made in Refs.~\cite{UO2-Gabi,UO2-Werner}, where it was observed that the $5f_{5/2}$ states
are Mott localized, while the $5f_{7/2}$ states are extended (but gapped).
However, these studies did not investigate how
this conclusion is influenced by the crystal field effects,
which is the main goal of this paper.
For this purpose, we perform charge self-consistent
LDA+RISB simulations of paramagnetic UO$_2$ taking fully into account
the CFS.
As in Ref.~\cite{Our-PRX}, we utilize
the density functional theory~\cite{HohenbergandKohn} 
code WIEN2K~\cite{w2k}
and employ the standard "fully localized limit"
form for the double-counting
functional~\cite{LDA+U}.
These calculations would have been prohibitive
without the algorithms derived in this work~\cite{supplemental_material}.

\begin{figure}
	\begin{center}
		\includegraphics[width=8.4cm]{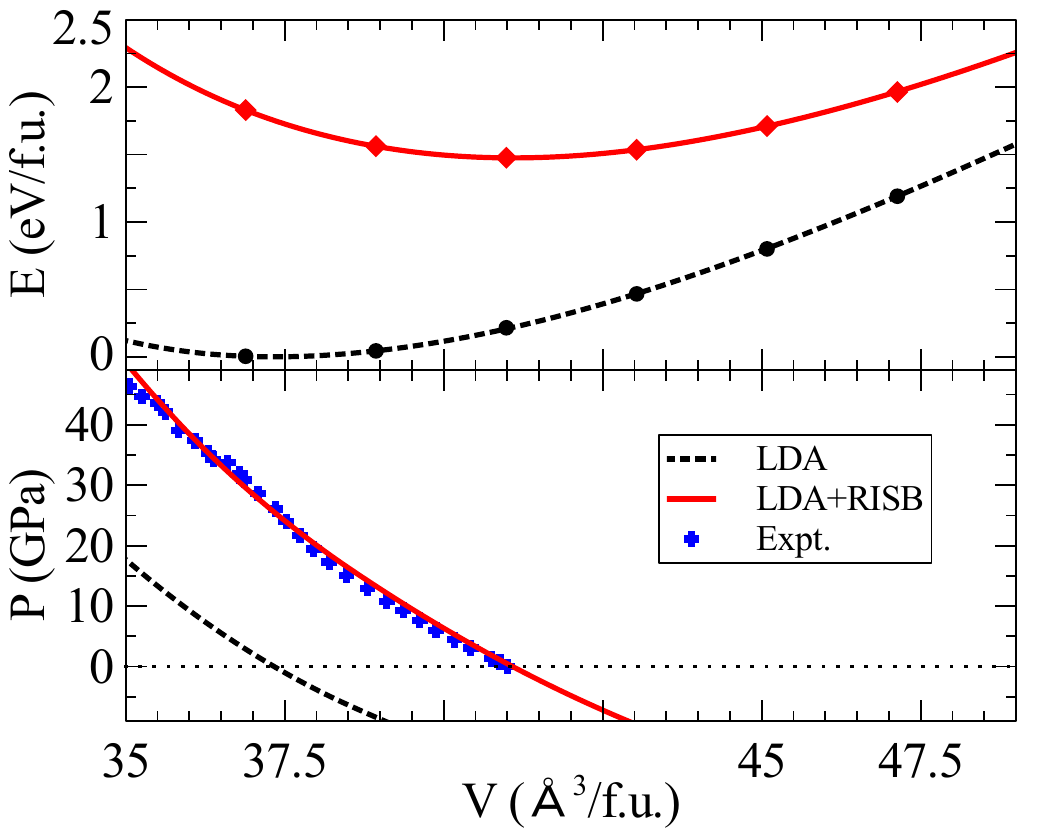}
		\caption{(Color online) 
			Zero temperature LDA and LDA+RISB total energies (upper panel)
			and corresponding pressure-volume phase diagrams
			compared with the room-temperature experiments of Ref.~\cite{UO2-exp}
			(lower panel).
		}
		\label{figure1}
	\end{center}
\end{figure}
As in Ref.~\cite{UO2-Werner}, in this work we assume that
the Hund's coupling constant is $J=0.6\,eV$.
In the upper panel of Fig.~\ref{figure1} are shown the LDA and LDA+RISB
total energies $E(V)$ obtained at zero temperature for
$U=10\,eV$~\cite{supplemental_material}.
%
The corresponding pressure (P-V) curves, obtained from $P(V)=-dE/dV$,
are shown in the lower panel in comparison with
the experimental data of Ref.~\cite{UO2-exp} (which were obtained
at room temperature).
The RISB P-V curve and, in particular, 
the experimental equilibrium volume
$V_{\text{eq}}\simeq 41\,\AA^3/\text{f.u.}$,
compare remarkably well with the experiments.
This favorable comparison with the experiments
gives us confidence that our theoretical approach is able to describe
the ground-state properties of this material.
As shown in the supplemental material~\cite{supplemental_material},
the P-V curve (and, in particular, the equilibrium volume) is essentially
identical for $U=8\,eV$, which is the value assumed in Ref.~\cite{UO2-Werner}.
Furthermore, reducing $U$ from $10\,eV$ to $8\,eV$
does not influence appreciably the electronic structure of UO$_2$ at $V_{\text{eq}}$~\footnote{Smaller values of $U$ have not been considered because,
	within our LDA+RISB functional, the 
	system would result metallic for 
	$U<6\,eV$ (which is the value of the screened Hubbard interaction
	parameter previously computed in Ref.~\cite{Amadon-UO2}) and, at the same
	time, the agreement with the experimental P-V curve would worsen.}.

In order to describe the orbital differentiation in UO$_2$
taking into account the CFS, it is necessary to decompose
the U-$5f$ single-particle space in irreducible representations
of the double $O$ point symmetry group~\cite{groups-Dresselhaus,Wigner} of the U atoms.
It can be shown that this repartition consists in:
1 $\Gamma_6(2)$ doublet, 2 $\Gamma_7(2)$ doublets and 2 $\Gamma_8(4)$ quartets~\footnote{In this work
	we adopted the so called Koster notation.}.
These irreducible representations are generated by the following states:
\bea
\vspace{-0.035cm}
\ket{\Gamma_6,7/2,\pm}\!&=&\!\sqrt{5/12}\,\ket{7/2,\pm 7/2}\!+\!\sqrt{7/12}\,\ket{7/2,\mp 1/2}\nonumber\\
\ket{\Gamma_7,7/2,\pm}\!&=&\!\mp\sqrt{3/4}\,\ket{7/2,\pm 5/2}\!\pm\!\sqrt{1/4}\,\ket{7/2,\mp 3/2}\nonumber\\
\ket{\Gamma_8^{(1)},7/2,\pm}\!&=&\!\pm\sqrt{7/12}\,\ket{7/2,\pm 7/2}\!\mp\!\sqrt{5/12}\,\ket{7/2,\mp 1/2}\nonumber\\
\ket{\Gamma_8^{(2)},7/2,\pm}\!&=&\!\mp\sqrt{1/4}\,\ket{7/2,\pm 5/2}\!\mp\!\sqrt{3/4}\,\ket{7/2,\mp 3/2}\nonumber\\
\ket{\Gamma_7,5/2,\pm}\!&=&\!\sqrt{5/6}\,\ket{5/2,\pm 3/2}\!-\!\sqrt{1/6}\,\ket{5/2,\mp 5/2}\nonumber\\
\ket{\Gamma_8^{(1)},5/2,\pm}\!&=&\!\sqrt{1/6}\,\ket{5/2,\pm 3/2}\!+\!\sqrt{5/6}\,\ket{5/2,\mp 5/2}\nonumber\\
\ket{\Gamma_8^{(2)},5/2,\pm}\!&=&\!\ket{5/2,\pm 1/2}
\,,
\vspace{-0.035cm}
\eea
which are expressed in terms of the conventional basis of eigenstates of the total angular momentum
(JJ basis).
By virtue of the Schur lemma~\cite{Wigner}, 
the entries of the U-$5f$ self energy $\Sigma(\omega)$ 
coupling states belonging to inequivalent irreducible
representations are equal to 0.
However, the total angular momentum $J^2$ is not a good quantum number,
as the matrix elements of $\Sigma(\omega)$ coupling the following states
are allowed: 
$\ket{\Gamma_7,5/2,\pm}$ with $\ket{\Gamma_7,7/2,\mp}$,
$\ket{\Gamma_8^{(1)},5/2,\pm}$ with $\ket{\Gamma_8^{(2)},7/2,\mp}$ and
$\ket{\Gamma_8^{(2)},5/2,\pm}$ with $\ket{\Gamma_8^{(2)},7/2,\pm}$.
Furthermore, the 5/2 and 7/2 states are not
degenerate~\cite{supplemental_material}.
Note that these CFS are present because of the crystal structure,
and would not exist if the environment of the $U$ atoms
was isotropic.

The main goals of this work are: (1) to show that
the CFS affect substantially the electronic structure 
of UO$_2$, and (2) to describe and explain the
pattern of orbital differentiation of the U-$5f$ electrons
in this material.

\begin{table}
	\begin{center}      
		\caption{Eigenvalues of the $5f$ quasi-particle matrix $Z$ and corresponding
			orbital occupations for LDA+RISB calculations at $U=10\,eV$.
			Theoretical results obtained by taking into account the crystal
			field splittings and by neglecting them.
		}
		\begin{ruledtabular}
			\begin{tabular}{llllll}
				w/ CFS     & $\Gamma_8(4)$ & $\Gamma_7(2)$ & $\Gamma_8(4)$ & $\Gamma_7(2)$ & $\Gamma_6(2)$ \\
				\hline
				$Z$   & $0$ & $0.92$ & $0.92$ & $0.95$ & $0.95$  \\
				\hline
				$n$   & $1.92$ & $0.14$ & $0.08$ & $0.06$ & $0.04$  \\
			\end{tabular}
			\begin{tabular}{lll}
				w/o CFS    & $5/2$ &  $7/2$   \\
				\hline
				$Z$        & $0$   &  $0.96$   \\
				\hline
				$n$        & $1.98$   &  $0.16$   \\
			\end{tabular}
			\label{table1}
		\end{ruledtabular}
	\end{center}
\end{table}
In Table~\ref{table1} are shown 
the eigenvalues of the $5f$ quasi-particle matrix
$Z=\R^\dagger\R$ obtained by taking into account the CFS
and the corresponding orbital occupations.
The approximate results calculated
by averaging over the CFS
are also shown.
The details of the averaging procedure
are described in the supplemental material. 
We observe that when the CFS are taken into account the selective
Mott localization occurs only within the $\Gamma_8$ sector, while the 
eigenvalues of $Z$ of the other $5f$ degrees of freedom are relatively large.
More precisely, $Z$ has 4 null eigenvalues with $\Gamma_8$ character.
On the other hand, when the CFS are neglected~\cite{UO2-Gabi,UO2-Werner}, 
the Mott localization can only occur
within the entire 5/2 sector, which is 6 times degenerate.
It is important also to observe that 
when the CFS are taken into account
the Mott localized $\Gamma_8$ states 
do not have a well defined total angular momentum $J^2$.
In fact, we found that the eigenstates of $Z$ with null eigenvalues
are the following:
\bea
\vspace{-0.035cm}
\ket{1} &\simeq& 0.939\, \ket{\Gamma_8^{(1)},5/2,+} + 0.343\, \ket{\Gamma_8^{(2)},7/2,-}\nonumber\\
\ket{2} &\simeq& 0.939\, \ket{\Gamma_8^{(1)},5/2,-} + 0.343\, \ket{\Gamma_8^{(2)},7/2,+}\nonumber\\
\ket{3} &\simeq& 0.939\, \ket{\Gamma_8^{(2)},5/2,+} + 0.343\, \ket{\Gamma_8^{(1)},7/2,-}\nonumber\\
\ket{4} &\simeq& 0.939\, \ket{\Gamma_8^{(2)},5/2,-} + 0.343\, \ket{\Gamma_8^{(1)},7/2,+}\,,
\vspace{-0.035cm}
\eea
which have considerably mixed $J^2$ character.
A further indication of the importance of the CFS in UO$_2$
is given by the orbital occupations of the U-$5f$ electrons.
In fact, the occupation corresponding to the Mott localized $5f$ electrons is
$1.92$, while the remaining $0.32$ $5f$ electrons are extended (but gapped).
Instead, when the CFS are neglected, the total number of Mott localized $5f$
electrons is $1.98$, while the occupation of the extended $5f$ degrees of freedom
is only $0.16$.
The fact that the overall occupancy of the $5f$ levels deviates considerably from
an integer value confirms the importance
of covalency effects in UO$_2$, which has been pointed out also in previous experimental
and theoretical studies~\cite{Covalency1,Covalency2,Covalency3,mixval_UO2_URu2Si2}.
Note also that the Mott-localized $\Gamma_8$ degrees of freedom 
have occupancy close to integer, which is a factor that
is known to promote localization~\cite{XiDai_Mott-IntegerOccupation}.

Let us now address the question of what is
the physical origin of the strong CFS orbital differentiation in UO$_2$.
The first important observation is that the importance of the CFS splittings
in UO$_2$ is not related with the U-$5f$ crystal fields
(on-site energy splittings)~\cite{XiDai_Mott-IntegerOccupation,orbital-selective-Mott-2,FeSe-FeTe},
which are very small in this material ($\sim 7\,meV$).
In fact, a direct calculation shows that neglecting the CFS
contributions to the on-site energy splittings~\cite{supplemental_material}
does not affect sensibly any of the results considered above (data not shown).
Furthermore, we find that the total energy of the approximate solution obtained by averaging
over the crystal fields is about $0.59\,eV/f.u.$ higher with respect to the
solution where the CFS are taken into account,
which is a much larger energy scale with respect to the 
above mentioned on-site energy splittings.
These observations and the data in Table~\ref{table1}
indicate that the main physical
reason why it is essential to take 
into account the CFS concerns
the above mentioned covalent nature 
of the bonds in UO$_2$, i.e.,
the hybridization between the U-$5f$ and the uncorrelated
electrons (in particular, the O-$2p$ states).
In particular, we note that neglecting the CFS
implies (by construction) that the $\ket{\Gamma_7,5/2,\pm}$ electrons
are Mott localized, which leads
to an underestimation of the contributions to the energy arising from the
hybridization of these electrons with the O-$2p$ bands.
On the other hand, taking into account the CFS enables to capture the fact
that the hybridization of the $\Gamma_7$ electrons is larger
with respect to the $\Gamma_8$ localized states~\cite{UO2-Savrasov}.

More details about the electronic structure
of UO$_2$ are reported in the supplemental material~\cite{supplemental_material}.

In summary, we have derived an exact
RISB reformulation of the multiband Hubbard model,
which establishes the foundation of the mean-field approximation and 
constitutes a starting point for calculations beyond mean-field.
The gauge invariance of our theory resulted also in substantial
algorithmic advancements, which make it possible to study from first principles the energetics and the electronic structure
of strongly correlated materials taking
into account simultaneously electron
correlations, SOC and CFS.
By utilizing
our theoretical approach, we have performed first principle
calculations of the orbital-selective Mott insulator UO$_2$, 
finding good agreement with available experimental data.
Furthermore, we have demonstrated
that taking into account the CFS is essential in order to capture the
correct pattern of orbital differentiation between the U-$5f$ states,
and that the main physical reason underlying
the CFS orbital differentiation in UO$_2$ is not the contribution of the
crystal field on-site energies (which is essentially negligible),
but concerns the hybridization between the U-$5f$
and the O-$2p$ electrons~\cite{UO2-Savrasov}, which originates
covalent bonds in this material~\cite{Covalency1,Covalency2,Covalency3,mixval_UO2_URu2Si2}.
The strong orbital differentiation between
the $\Gamma_8$ and the $\Gamma_7$ electrons
could be directly detected
experimentally, e.g., by means of angle-resolved photoemission
techniques~\cite{angle-resolved-photoemission-1,angle-resolved-photoemission-2}, 
which would enable us 
to discriminate between the spectral contributions of the different states
based on their symmetry properties.
In particular,
based on the orbital occupations of Table~\ref{table1}
and the Friedel sum rule, we predict that
the $5f$ spectral weight~\cite{UO2-spectrum1,UO2-spectrum2} below the Fermi level 
has mostly $\Gamma_8$ character --- while it would have also
a substantial $\Gamma_7$ contribution 
if the CFS orbital differentiation was a negligible effect.
The analysis presented here is very general and could be applied 
also to other $f$ electron systems, 
e.g., to materials displaying strong magnetic anisotropy
or more general forms of multipolar order~\cite{RevModPhys-Actinides_oxides}.

\begin{acknowledgments}
	We thank Cai-Zhuang Wang, Kai-Ming~Ho and Tsung Han for useful discussions.
	This research was supported by the U.S. Department of energy, Office of Science,
	Basic Energy Sciences, as a part of the Computational Materials Science Program. 
	V.D. and N.L. were partially supported by the NSF grant DMR-1410132
	and the National High Magnetic Field Laboratory.
\end{acknowledgments}

N.L. and Y.Y. equally contributed to this work.
N.L. contributed mostly to the formal and algorithmic aspects of the theory
and Y.Y. contributed mostly to the numerical implementation.
X.D. performed part of the calculations of UO$_2$.
All the authors contributed to write the manuscript.
G.K. supervised the project.


%

\widetext
\clearpage

\begin{center}
\textbf{\large Supplemental Material:\\ Operatorial Formulation of the Rotationally Invariant Slave Boson Theory and Mapping between Slave Boson Amplitudes and Embedding System}
\end{center}

\renewcommand\thesection{\arabic{section}}
\setcounter{equation}{0}
\setcounter{figure}{0}
\setcounter{table}{0}
\setcounter{page}{1}
\makeatletter

\begin{changemargin}{1.5cm}{1.5cm} 
In this supplemental material we provide the details of the
construction of the RISB renormalization operators.
Furthermore, we discuss the most important technical and algorithmic
advantages of the gauge invariance formulation of the RISB mean field
theory presented in the main text with respect to the formulation of
Ref.~\onlinecite{Our-PRX-supplemental}.
Finally, we present several additional details about our calculations of UO$_2$.
In particular, we explain the exact definition
of the averaging procedure with respect to the crystal field splittings,
which was introduced in the main text.
Furthermore, we present a few additional details
about the electronic structure of this material.
\end{changemargin}

\section{I.~~~Construction of the RISB Hamiltonian}

In the main text we have defined the physical subspace $h_{\text{SB}}$ 
as the subspace of the RISB Hilbert space $\mathcal{H}_{\text{SB}}$ 
satisfying the following equations, which are called ``Gutzwiller constraints'':
\bea
\sum_{An}\Phi^\dagger_{RiAn}\!\Phi^\dagga_{RiAn} \!&=&\! 1 
\;\;\forall\,R,i
\label{C1-supplemental}
\\
\sum_{Anm} [F^\dagger_{ia}F^\dagga_{ib}]_{mn}\,
\Phi^\dagger_{Ri An}\!\Phi^\dagga_{Ri Am}
\!&=&\! \fc_{Ria}\fa_{Rib}\;\;\forall\, R,i,a,b
\,,~~~~~
\label{C2-supplemental}
\eea
where
\be
\left[F_{ia}\right]_{nm}\equiv
\langle n,Ri|\,\fa_{Ria}\,| m,Ri\rangle\,.
\ee
In Ref.~\onlinecite{Georges-supplemental} it was shown that
$h_{\text{SB}}$ is spanned by the following states:
\bea
\uA &=& \frac{1}{\sqrt{D_{iA}}}\,\sum_n\Phi^\dagger_{Ri An}
\big[f^\dagger_{Ri 1}\big]^{\nu_{1}(n)}\!\!\!\!\!\!.\,.\,.\;
\big[f^\dagger_{Ri M_i}\big]^{\nu_{M_i}(n)} |0\rangle
= \mathcal{U}\,\ket{A,Ri}
\,,
\label{phys_SB-supplemental}
\eea
where $D_{iA}\equiv\binom{M_i}{N_A}$ is a binomial coefficient,
which enforces the normalization of these states.
In fact, it can be readily verified that:
\be
\langle\underline{A},Ri | \underline{B},R'j\rangle = 
\langle A,Ri | B,R'j\rangle =
\delta_{RR'}\delta_{ij}\delta_{AB}\,.
\ee
The unitary operator $\mathcal{U}$ defined in Eq.~\eqref{phys_SB-supplemental}
defines the mapping between the original Fock space and $h_{\text{SB}}$.

\subsection{A.~~~The RISB Renormalization Operators}

In this subsection we will construct explicitly
the RISB renormalization operators $\Rh_{Ri a\alpha}$
introduced in the main text.
Our goal consists in constructing with
$\{\Phi^\dagga_{RiAn}\}$ and $\{\Phi^\dagger_{RiAn}\}$
a set of operators $\Rh_{Ria\alpha}$
such that the operators
\be
\ccu_{Ri\alpha}\equiv
\sum_a\Rh_{Ria\alpha}[\Phi^\dagga_{RiAn},\Phi^\dagger_{RiAn}]
\,\fc_{Ri a}
\label{c_underlined_R-supplemental}
\ee
satisfy the following property:
\be
\langle \underline{A}, Ri |\, \ccu_{Ri\alpha} \,| \underline{B}, Ri \rangle
= \langle A, Ri |\, \cc_{Ri\alpha} \,| B, Ri \rangle
\quad \forall\,A,B\,.
\label{c_underlined_property-supplemental}
\ee
Furthermore, we require that our renormalization operators
reproduce the mean field equations of Ref.~\onlinecite{Georges-supplemental}.

We will proceed by providing directly the operators $\Rh_{Ria\alpha}$ and
demonstrating that they satisfy the above mentioned requirements
by inspection.

Let us introduce the matrices:
\bea
\Deltap_{Ri ab} &\equiv& \sum_{Anm} [F^\dagger_{ia}F^\dagga_{ib}]_{mn}\,
\Phi^\dagger_{Ri An}\Phi^\dagga_{Ri Am}\\
\Deltah_{Ri ab} &\equiv& \sum_{Anm} [F^\dagga_{ib}F^\dagger_{ia}]_{mn}\,
\Phi^\dagger_{Ri An}\Phi^\dagga_{Ri Am}
\,.
\eea 
Note that the elements $(a,b)$ of $\Dp$ and $\Dh$ are operators.
For later convenience, we define also the corresponding 
operatorial matrix products:
\bea
[\Dp\bullet\Dh]_{Ri ab}\equiv\Deltap_{Ri ac}\Deltah_{Ri cb}
\eea
and the powers:
\bea
[\Dp]^{\left[l\right]}_{Ri ab} &\equiv&
\Deltap_{Ri ac_1}\Deltap_{Ri c_1c_2}\,...\,\Deltap_{Ri c_{l-1}b}\\~
[\Dh]^{\left[l\right]}_{Ri ab} &\equiv&
\Deltah_{Ri ac_1}\Deltah_{Ri c_1c_2}\,...\,\Deltah_{Ri c_{l-1}b}\\~
[\hat{\Delta}_p]^{[l=0]}_{Ri ab}
&=&[\hat{\Delta}_h]^{[l=0]}_{Ri ab} \equiv \delta_{ab}
\,,
\eea
where the symbols ``$[l]$'' and ``$\bullet$''
indicate that we are doing matrix products.
Finally, we introduce the following
series of operators:
\bea
\left[\hat{1}-\Dp\right]^{[-\frac{1}{2}]} &\equiv&
\sum_{r=0}^\infty (-1)^r \binom{\frac{1}{2}}{r}\,\Deltap^{[r]}
\label{Dp-op-supplemental}
\\
\left[\hat{1}-\Dh\right]^{[-\frac{1}{2}]} &\equiv&
\sum_{r=0}^\infty (-1)^r \binom{\frac{1}{2}}{r}\,\Deltah^{[r]}
\label{Dh-op-supplemental}
\,,
\eea
where $\binom{a}{b}$ is the usual
notation for the binomial coefficient and $\hat{1}$ indicates the identity operator.

As we are going to show below, the following renormalization operators
satisfy the desired properties, i.e.,
Eqs.~\eqref{c_underlined_R-supplemental} and \eqref{c_underlined_property-supplemental}:
\bea
\Rh_{Ri a\alpha}&\equiv&\sum_{ABnmb}
\frac{[F^\dagger_{i\alpha}]_{AB}[F^\dagger_{ib}]_{nm}}{\sqrt{N_A(M_i\!-\!N_B)}}
\nonumber\\&&\qquad 
:\Phi^\dagger_{Ri An}\!\left[
\hat{1}\!+\!\left(\!\sqrt{N_A(M_i\!-\!N_B)}\!-\!1\!\right)\!
\sum_{Cl}\Phi^\dagger_{Ri Cl}\Phi^\dagga_{Ri Cl}
\right]\!\!
\left[
\left[\hat{1}-\Dp\right]^{[-\frac{1}{2}]}\!\!\!\bullet\!\left[\hat{1}-\Dh\right]^{[-\frac{1}{2}]}
\right]_{Ri ba}\!\!\!
\Phi^\dagga_{Ri Bm}:\,,
\label{Bosonic_R-supplemental}
\eea
where ``$:$'' indicates the normal ordering.

Note that Eq.~\eqref{Bosonic_R-supplemental} contains a term proportional to
$\sum_{Cl}\Phi^\dagger_{Ri Cl}\Phi^\dagga_{Ri Cl}$, which
was not present in the definition of Ref.~\onlinecite{Georges-supplemental}.
It is thanks to this additional term that, as we are going to show,
Eq.~\eqref{Bosonic_R-supplemental} reproduces the GA
at the mean-field level while
it is --- at the same time ---
also fully justified from the operatorial perspective.

\subsubsection{1.~~~Proof that $\Rh_{Ri ab}$ have correct action on physical states}

In order to prove that $\Rh_{Ri ab}$ satisfies
Eqs.~\eqref{c_underlined_R-supplemental} and \eqref{c_underlined_property-supplemental}
we observe that these operators act on the physical states 
exactly as
\bea
&&\hat{R}_{Ri a\alpha}\equiv\sum_{AB}\sum_{nm}
\frac{[F^\dagger_{i \alpha}]_{AB}[F^\dagger_{i a}]_{nm}}{\sqrt{N_A(M_i-N_B)}}\,
\Phi^\dagger_{Ri An} \Phi^\dagga_{Ri Bm}
\label{old_R-supplemental}\\
&&\;=\sum_{AB}\sum_{nm}\frac{1}{N_A}\sqrt{\frac{D_{iB}}{D_{iA}}}\,
[F^\dagger_{i \alpha}]_{AB}[F^\dagger_{i a}]_{nm}
\Phi^\dagger_{Ri An} \Phi^\dagga_{Ri Bm}
\,,
\nonumber
\eea
see Eq.~\eqref{phys_SB-supplemental},
which were shown to have the correct action over the physical space
in Ref.~\onlinecite{Georges-supplemental}.

As discussed in the main text,
the reason why Eqs.~\eqref{Bosonic_R-supplemental} and \eqref{old_R-supplemental}
are equivalent within the subspace of physical sates is that,
since the bosonic operators are normally ordered,
all of the terms of Eq.~\eqref{Bosonic_R-supplemental}
containing more than one bosonic annihilation operator
are zero when they act on the physical states, see Eq.~\eqref{phys_SB-supplemental}.

It is useful to observe that, thanks to the normal ordering,
Eq.~\eqref{Bosonic_R-supplemental} is well defined
not only within the subspace of physical states, but 
also on the states with any finite number of bosonic operators.
In fact, if Eq.~\eqref{Bosonic_R-supplemental} is applied to any
state with $n_B$ slave bosons (or less),
the terms of the series [Eqs.~\eqref{Dp-op-supplemental} and \eqref{Dh-op-supplemental}]
with $r>n_B$ do not contribute.

\subsubsection{2.~~~Mean field renormalization factors}

Let us now prove that Eq.~\eqref{Bosonic_R-supplemental} reproduces the 
renormalization coefficients of Ref.~\onlinecite{Georges-supplemental} at the mean-field level.

As discussed in the main text,
the zero-temperature RISB mean-field theory consists in searching
the ground state of the $\hu$ in the whole RISB Hilbert space
assuming a variational wavefunction represented as
\be
\ket{\Psi_{\text{SB}}} = \ket{\Psi_0}\otimes\ket{\phi}\,,
\label{varsb-supplemental}
\ee
where $\ket{\Psi_0}$ is a Slater determinant constructed with
the quasi-particle ladder operators $\fa_{Ria}$, $\ket{\phi}$
is a bosonic coherent state, and the Gutzwiller constraints,
see Eqs.~\eqref{C1-supplemental} and \eqref{C2-supplemental}, are enforced only in average.

It can be verified that taking the expectation value of 
Eqs.~\eqref{C1-supplemental} and \eqref{C2-supplemental} with respect to the variational
state [Eq.~\eqref{varsb-supplemental}] gives the following equations:
\bea
\Tr\!\left[\phi^\dagger_{i}\phi^\dagga_{i}\right] \!&=&\! 1 
\;\forall\,i
\label{avC1-supplemental}
\\
\Tr\!\left[\phi^\dagger_{i}\phi^\dagga_{i}\,F^\dagger_{ia}F^\dagga_{ib}\right]
\!&=&\! \Av{\Psi_0}{\fc_{Ria}\fa_{Rib}}\;\forall\, i,a,b\,,
\label{avC2-supplemental}
\eea
where the matrix elements $[\phi_i]_{An}$ are the eigenvalues of the ladder
operators $\Phi_{RiAn}$ with respect to the variational
coherent state $\ket{\phi}$.

Let us now calculate the average of Eq.~\eqref{Bosonic_R-supplemental} with respect
to a bosonic coherent state $\ket{\phi}$.
The essential observation
is that the term $\sum_{Cl}\Phi^\dagger_{Ri Cl}\Phi^\dagga_{Ri Cl}$ 
of Eq.~\eqref{Bosonic_R-supplemental} is equivalent to the identity at the
mean field level because of the first Gutzwiller constraint,
see Eq.~\eqref{avC1-supplemental}.
Consequently, this term cancels out the factors $\sqrt{N_A(M_i-N_B)}$
from Eq.~\eqref{Bosonic_R-supplemental}.
Thus, it can be straightforwardly verified that:
\bea
&&\R_{i a\alpha}[\phi] \equiv
\Av{\phi}{\Rh_{Ri a\alpha}}
\nonumber\\
&&\;=
\Tr\!\left[\phi_i^\dagger F^\dagger_{i\alpha}\phi^\dagga_i F^\dagga_{ib}\right]
\left[(1-[\Delta_{p i}])(1-[\Delta_{h i}])\right]^{-\frac{1}{2}}_{ba}
\nonumber\\
&&\;=
\Tr\!\left[\phi_i^\dagger F^\dagger_{i\alpha}\phi^\dagga_i F^\dagga_{ib}\right]
\left[
\Delta_{p i}(1-[\Delta_{p i}])\right]^{-\frac{1}{2}}_{ba}
\,,
\label{def-R-mf-supplemental}
\eea
where $1$ is the identity matrix ($1_{ab}=\delta_{ab}\,\forall a,b$), and
\bea
\left[\Delta_{pi}\right]_{ab} &\equiv& \Av{\phi}{\Deltap_{Ri ab}\!}=
\Tr\!\left[\phi^\dagger_i\phi^\dagga_i\,F^\dagger_{ia}F^\dagga_{ib}\right]
\label{def-Deltap-supplemental}\\
\left[\Delta_{hi}\right]_{ab} &\equiv& \Av{\phi}{\Deltah_{Ri ab}\!}=
\Tr\!\left[\phi_i^\dagger\phi^\dagga_i\,F^\dagga_{ib}F^\dagger_{ia}\right]
\eea 
are matrices of complex numbers.
Equation~\eqref{def-R-mf-supplemental} coincides with the mean field renormalization
matrices proposed in Ref.~\onlinecite{Georges-supplemental}.

\section{II.~~~Gauge Invariance RISB Hamiltonian: Proof of Eq.~8 main text}


\subsection{A.~~~Gauge group}

From Eqs.~\eqref{C1-supplemental} and \eqref{C2-supplemental} it follows that
\bea
\G^0(\zeta) &\equiv& e^{i\sum_{Ri}\zeta_{Ri}K^0_{Ri}}=1\quad\forall\,\zeta
\label{alternative-C1-supplemental}
\\
\G(\theta) &\equiv& e^{i\sum_{Riab}\theta_{Riab}K_{Riab}} = 1 \quad\forall\,\theta=\theta^\dagger
\label{alternative-C2-supplemental}
\eea
where
\bea
K^0_{Ri}&\equiv& \sum_{An} \Phi^\dagger_{Ri An}\!\Phi^\dagga_{Ri An} - I\\
K_{Riab}&\equiv&
\label{gauge-generators-supplemental}
\fc_{Ria}\fa_{Rib} - 
\sum_{Anm} [F^\dagger_{ia}F^\dagga_{ib}]_{mn}\,
\Phi^\dagger_{Ri An}\!\Phi^\dagga_{Ri Am}\,,~~~~
\eea
and $I$ is the identity operator.
We observe that:
\bea
&&\G(\theta)\, \Phi^\dagga_{RiAn} \,\G^\dagger(\theta)
\!=\!
\sum_m U_{Ri}(\theta_{Ri})_{mn}\Phi^\dagga_{RiAm}
\label{operatorial-gauge-bosonic-supplemental}\\
&&\G(\theta)\, \fc_{Ria}\, \G^\dagger(\theta)
=u_{Ri}(\theta_{Ri})_{ba}\,\fc_{Rib}
\label{operatorial-gauge-fermionic2-supplemental}
\,,
\eea
where
\be
U_{Ri}(\theta_{Ri})\equiv e^{i\sum_{ab}\theta_{Riab}F^\dagger_{ia}F^\dagga_{ib}}
\ee
and
\be
u_{Ri}(\theta_{Ri})\equiv e^{i\theta_{Ri}}
\ee
is the corresponding restriction within the
single-particle space.

\subsection{B.~~~The RISB Hamiltonian}

It can be readily verified that, as shown in Ref.~\onlinecite{Georges-supplemental}, the 
bosonic operator
\be
\hu^{\text{loc}} \equiv \sum_{Ri}\sum_{AB}
[H_i^{\text{loc}}]_{AB}\sum_n\Phi^\dagger_{RiAn}\Phi^\dagga_{RiBn}
\label{hloc_bosons-supplemental}
\ee
is a faithful representation of $\h^{\text{loc}}$,
i.e., that:
\be
\langle \underline{A}, Ri |\, \underline{\h}^{\text{loc}} \,| \underline{B}, Ri \rangle
= \langle A, Ri |\, \h^{\text{loc}} \,| B, Ri \rangle
\quad \forall\,A,B\,.
\label{hloc_reformulation-supplemental}
\ee

In summary, we have shown that the Hubbard Hamiltonian can be
equivalently represented in the RISB physical Hilbert space as follows:
\be
\hu=
\sum_{kij,\alpha\beta}\epsilon^{\alpha\beta}_{k,ij}\, \ccu_{ki\alpha}\cau_{kj\beta}
+\hu^{\text{loc}}\,,
\label{hu-supplemental}
\ee
where $\ccu_{ki\alpha}$ are the representation in momentum space
of the operators defined by
Eqs.~\eqref{c_underlined_R-supplemental} and \eqref{Bosonic_R-supplemental},
and $\hu^{\text{loc}}$ is given by Eq.~\eqref{hloc_bosons-supplemental}.

\subsection{C.~~~Gauge Invariance of RISB Hamiltonian} 

A remarkable property of $\hu$, see Eq.~\eqref{hu-supplemental} is that 
it is gauge invariant in the whole RISB Fock space $\mathcal{H}_{\text{SB}}$,
and not only within the subspace $h_{\text{SB}}$ of physical states.
In fact, it is straightforward to verify that
\bea
\G(\theta)\,
[\hat{\Delta}_{p}]_{Ri ab}\,
\G^\dagger(\theta)
&=& ^tu_i(\theta_i)_{aa'}
[\hat{\Delta}_{p}]_{Ri a'b'}
{^tu_i^\dagger}(\theta_i)_{b'b}~~~
\label{op-g-delta-supplemental}
\\
\G(\theta)\, \hat{R}_{Ri a\alpha}\, \G^\dagger(\theta)
&=& u^\dagger_i(\theta_i)_{ab}\,\hat{R}_{Ri b\alpha}
\label{op-g-R-supplemental}
\\
\G(\theta)\, \fc_{Ria}\, \G^\dagger(\theta)
&=&u_i(\theta_i)_{ba}\,\fc_{Rib}\\
\G(\theta)\, \hu^{\text{loc}}\, \G^\dagger(\theta)
&=&\hu^{\text{loc}}\,,
\eea
and that, consequently,
\be
\G(\theta)\, \hu\, \G^\dagger(\theta)
=\hu\quad\forall\,\theta=\theta^\dagger
\label{operatorial-gauge-invariance-supplemental}
\,.
\ee
This completes the proof of Eq.~8 of the main text.

\section{III.~~~The RISB mean-field Lagrange function} 

Let us consider the RISB theory at the mean field level, which was introduced
in the main text.
Similarly to Ref.~\onlinecite{Our-PRX-supplemental},
the corresponding energy constrained minization problem
can be conveniently formulated by utilizing the
following Lagrange function:
\vspace{-0.12cm}
\bea
&&\Lag_{\text{SB}}[\phi,E^c;\,  \R,\R^\dagger,\lambda;\, \D,\D^\dagger, \lambda^{c};\,\Delta_p]
= -\lim_{\mathcal{T}\rightarrow 0}\frac{\mathcal{T}}{\mathcal{N}}\sum_{k}\sum_{m\in\mathbb{Z}}
\Tr\log\!
\left(\frac{1}{i (2m+1)\pi\mathcal{T}-\R\epsilon_{k}\R^\dagger-\lambda+\mu}\right)
e^{i(2m+1)\pi\mathcal{T} 0^+}
\nonumber\\[-0.6mm]
&&+\sum_i\Tr
\bigg[\phi_i^\dagga\phi_i^\dagger\,H^{\text{loc}}_i
\!+\!\sum_{a\alpha} \left(
\left[\D_{i}\right]_{a\alpha}
\,\phi_i^\dagger\,F^\dagger_{i\alpha}\,\phi^\dagga_i\,F^\dagga_{ia}+\text{H.c.}\right)
\!+\!\sum_{ab} \left[\lambda^c_{i}\right]_{ab}\,\phi_i^\dagger\phi^\dagga_i\,F^\dagger_{ia}F^\dagga_{ib}\bigg] 
\!+\!\sum_iE^c_i\!\left(1\!-\Tr\big[\phi_i^\dagger\phi^\dagga_i\big]\right)
\nonumber\\[-0.6mm]
&&-\sum_i\bigg[
\sum_{ab}\big(
\left[\lambda_i\right]_{ab}+\left[\lambda^c_i\right]_{ab}\big)\left[\Delta_{pi}\right]_{ab}
+\sum_{c a\alpha}\left(
\left[\D_{i}\right]_{a\alpha}\left[\R_{i}\right]_{c\alpha}
\big[\Delta_{pi}(1-\Delta_{pi})\big]^{\frac{1}{2}}_{c\alpha}
+\text{c.c.}\right)
\bigg]\,.
\label{Lag-SB-supplemental} 
\eea
\vspace{-0.12cm}

As in Ref.~\onlinecite{Our-PRX-supplemental}, $\lambda_i^c$, $\lambda_i$ and $\D_i$ are
matrices of Lagrange multipliers:
(i) $\lambda^c_i$ enforces
the definition of $\Delta_{pi}$ in terms of the RISB amplitudes,
see Eq.~\eqref{avC2} (left);
(ii) $\lambda_i$ enforces the Gutzwiller constraints,
see Eq.~\eqref{avC2} (right); and
(iii) $\D_i$ enforces the definition of $\R_i$,
see Eq.~\eqref{def-R-mf}.
The main advantage of this reformulation is that $\Lag_{\text{SB}}$
depends only \emph{quadratically} on the RISB amplitudes.

\subsection{A.~~~Gauge transformation}

It can be readily verified by inspection that $\Lag_{\text{SB}}$
is invariant with respect to the following
group of gauge transformations:
\bea
\phi_i &\longrightarrow&\phi_i\,U_i(\theta_i)\,,\;\,
\Delta_{pi} \longrightarrow {^tu_i}(\theta_i)\,\Delta_{pi}\,{^tu_i}^\dagger(\theta_i)\,,
\label{gauge-Deltap-supplemental}\\[-0.5mm]
\R_i &\longrightarrow& u_i^\dagger(\theta_i)\,\R_i\,,\;\,
\lambda_i \longrightarrow u_i^\dagger(\theta_i)\,\lambda_i\,u_i(\theta_i)\,,
\label{gauge-R-supplemental}\\[-0.5mm]
\D_i &\longrightarrow& {^tu_i}(\theta_i)\,\D_i\,,\;\,
\lambda^c_i \longrightarrow u_i^\dagger(\theta_i)\,\lambda_i^c\,u_i(\theta_i)
\,,
\eea
where $U_{i}(\theta_{i})\equiv e^{i\sum_{ab}[\theta_{i}]_{ab}F^\dagger_{ia}F^\dagga_{ib}}$,
and $u_{i}(\theta_{i})\equiv e^{i\theta_{i}}$
is the corresponding restriction within the
single-particle space.
Consequently, given any set of RISB parameters
such that $\Lag_{\text{SB}}$ is stationary with respect to all of its arguments,
a manifold of infinite physically-equivalent
solutions can be found by applying to it
the above-mentioned continue group of Gauge transformations.

In order to study real materials it is often important to exploit the
point symmetry of the system, which enables us to reduce the dimensionality of the
manifold of RISB solutions, thus reducing the computational complexity of
the problem.
In particular, as we are goint to discuss, it is often
useful to transform a solution found in a given basis into a different
representation.
For this purpose, it is desirable to work with a Lagrange function
which is explicitly covariant with respect to the point group of the
system.

In this section we are going to show that while the gauge-invariant
Lagrange function is explicitly covariant under changes of basis with
respect to the symmetry point group of the system, the natural-basis
gauge fixing breaks this property (as it happens in electrodynamics).

\subsection{B.~~~Change of basis} 

Let us assume that we have found a saddle point of the RISB Lagrange function
in a given basis, so that the dispersion is $\epsilon_{k,ij}$ and
the coefficients appearing in Eq.~\eqref{hloc_bosons-supplemental}
are the elements of a given set of matrices $H_i^{\text{loc}}$.
Then, we reformulate the same problem in a new basis obtained from the
previous by applying the following local change of basis:
\be
\cc_{Ri\alpha}\longrightarrow
\bar{L}^\dagga_{Ri}\,\cc_{Ri\alpha}\,\bar{L}_{Ri}^\dagger
\equiv \sum_{\alpha'}[L_{i}]_{\alpha'\alpha}\,\cc_{Ri\alpha'}\,,
\label{change-basis-supplemental}
\ee
so that
\bea
\epsilon_{k,ij}&\longrightarrow&
L^\dagger_{i}\,\epsilon_{k,ij}\,L_{j}^\dagga\\
H_i^{\text{loc}}&\longrightarrow&
\bar{L}^\dagger_{Ri}\,H_i^{\text{loc}}\,\bar{L}_{Ri}^\dagga
\label{tb-Hloc-supplemental}
\,.
\eea

It can be readily verified that, within the gauge invariant Lagrange
formulation, the RISB solution transforms as follows under the above-mentioned
change of basis:
\bea
\phi_i &\longrightarrow& \bar{L}^\dagger_{Ri}\,\phi_i\,\bar{L}^\dagga_{Ri}
\label{tb1-supplemental}\\
\Delta_{pi} &\longrightarrow& ^tL^\dagga_i\,\Delta_{pi}\,^tL_i^\dagger
\label{tb2-supplemental}\\
\lambda^c_i &\longrightarrow& L_i^\dagger\,\lambda_i^c\,L^\dagga_i
\label{tb3-supplemental}\\
\lambda_i &\longrightarrow& L_i^\dagger\,\lambda_i\,L^\dagga_i
\label{tb4-supplemental}\\
\R_i &\longrightarrow& L_i^\dagger\,\R_i\,L_i^\dagga
\label{tb5-supplemental}\\
\D_i &\longrightarrow& ^tL_i^\dagga\,\D_i\,^tL_i^\dagger
\label{tb6-supplemental}
\,.
\eea

Note that if the problem is formulated applying
the natural-basis gauge fixing
the transformations of the RISB variational parameters
are no longer similarity transformations.
For instance, it can be readily shown that:
\bea
\phi_i &\longrightarrow& \bar{L}^\dagger_{Ri}\,\phi_i\\
n^0 &\longrightarrow& n^0\\
\R_i &\longrightarrow& \R_i\,L_{Ri}
\,.
\eea

\subsection{C.~~~Imposing the symmetries} 

Let us assume that the Hubbard Hamiltonian is expressed in a given
basis $\cc_{Ri\alpha}$, and that the system is invariant
with respect to a given point group
$\{\bar{g}_{Rin}\}\equiv\bar{G}_{Ri}$ of symmetry transformations
centered at the site $(R,i)$
such that the ladder operators transform as follows:
\be
\cc_{Ri\alpha}\longrightarrow
\bar{g}^\dagga_{Rin}\,\cc_{Ri\alpha}\,\bar{g}^\dagger_{Rin}
=\sum_{\alpha'}[g_{Rin}]_{\alpha'\alpha}\,\cc_{Ri\alpha'}
\label{sp-g-supplemental}
\,.
\ee

In order to exploit the symmetry defined above
it is convenient to choose a basis such that the matrices $g_{Rin}$
are represented as a sum of irreducible representations and these
representations are set to be equal whenever they are equivalent.
From now on we are going to define such a basis a ``symmetry basis''.
A practical method to construct such a representation is provided in the
supplemental material.

As shown in Refs.~\onlinecite{Gmethod-supplemental}, if the Hubbard Hamiltonian is represented
in a symmetry basis, the condition that both the Gutzwiller projector
and the GA variational Slater determinant are invariant with respect
to $\bar{G}_{Ri}$ amounts to impose that the RISB amplitudes satisfy
the following condition:
\be
[\bar{g}_{Rin},\phi_i]=0\quad\forall\,\bar{g}_{Rin}\in \bar{G}_{Ri}\,.
\label{sym-phi-supplemental}
\ee
This condition reduces the dimension of the most general
matrix $\phi_i$ respecting the symmetries in the way established by
the Shur lemma.

From the definitions of $\Delta_{pi}$ and $\R_{i}$,
see Eqs.~\eqref{def-Deltap-supplemental} and ~\eqref{def-R-mf-supplemental},
and from Eq.~\eqref{sym-phi-supplemental} it can be readily verified that
\be
[^tg_{Rin},\Delta_{pi}]=[g_{Rin},\R_{i}]=
0\quad\forall\,\bar{g}_{Rin}\in \bar{G}_{Ri}
\label{sym-sp-1-supplemental}
\,,
\ee
where the single-particle matrices $g_{Rin}$ were defined
in Eq.~\eqref{sp-g-supplemental}.
Since $\D_i$, $\lambda_i$ and $\lambda^c_i$ are matrices
of Lagrange multipliers, they retain the structure of their
conjugate variables. Consequently, they satisfy the
following relations:
\be
[^tg_{Rin},\D_{pi}]=[g_{Rin},\lambda_{i}]=[g_{Rin},\lambda^c_{i}]=
0\quad\forall\,\bar{g}_{Rin}\in \bar{G}_{Ri}
\label{sym-sp-2-supplemental}
\,.
\ee

We point out that working with the gauge-invariant Lagrange function,
see Eq.~\eqref{Lag-SB-supplemental}, has the advantage that in this formulation
the symmetry conditions on the variational parameters 
are covariant with respect to changes of basis,
i.e.: 
\bea
\left[\bar{g}_{Rin},\phi_i\right]=0 & \implies & \left[\bar{g}'_{Rin},\phi'_i\right]=0
\label{gtb1-supplemental}\\
\left[^tg_{Rin},\Delta_{pi}\right]=0  &\implies& \left[^tg'_{Rin},\Delta'_{pi}\right]=0
\label{gtb2-supplemental}\\
\left[g_{Rin},\lambda^c_{i}\right]=0 & \implies &  \left[g'_{Rin},\lambda^{c\prime}_{i}\right]=0
\label{gtb3-supplemental}\\
\left[g_{Rin},\lambda_{i}\right]=0 & \implies &  \left[g'_{Rin},\lambda'_{i}\right]=0
\label{gtb4-supplemental}\\
\left[g_{Rin},\R_{i}\right]=0 & \implies &  \left[g'_{Rin},\R'_{i}\right]=0
\label{gtb5-supplemental}\\
\left[^tg_{Rin},\D_{i}\right]=0  &\implies& \left[^tg'_{Rin},\D'_{i}\right]=0
\label{gtb6-supplemental}
\,,
\eea
where $\phi'_i$, $\Delta'_{pi}$, $\lambda^{c\prime}_{i}$, $\lambda'_{i}$, $\R'_{i}$
and $\D'_{i}$ are the transformed of the RISB variational parameters
according to Eqs.~\eqref{tb1-supplemental}-\eqref{tb6-supplemental}, and
\bea
\bar{g}'_{Rin} &\equiv& \bar{L}^\dagger_{Ri}\,\bar{g}_{Rin}\,\bar{L}^\dagga_{Ri}\\
g'_{Rin} &\equiv& L^\dagger_{i}\,g_{Rin}\,L^\dagga_{i}
\,.
\eea

As we are going to see,
working with a Lagrange function explicitly covariant
under changes of basis turns out to be practically useful when the system
under consideration is constituted by a main term with high symmetry and a
smaller perturbation breaking part of its symmetry (which is a very common
situation).

\section{IV.~~~Reformulation using Embedding Hamiltonian}

In Ref.~\onlinecite{Our-PRX-supplemental} it was introduced a mapping between the matrices $\phi_i$
and the Hilbert space of states $\ket{\Phi_i}$ of an impurity system
composed by the $i$-impurity and an uncorrelated bath with the same
dimension, which provided an insightful physical interpretation
of the parameters $\phi_i$ based on the Schmidt decomposition.
In this section we will discuss this mapping in relation with the
transformation properties of the RISB solution under changes of basis
discussed in Sec.~III~B. 

For completeness, we first summarize the derivation of
the above-mentioned mapping.
Let us define a copy of the Fock space generated by the states
defined in Eq.~\eqref{phys_SB-supplemental}:
\bea
|A,i\rangle &\equiv& \big[\hat{c}^\dagger_{i 1}\big]^{\nu_{1}(A)}\!\!\!\!\!\!.\,.\,.\;
\big[\hat{c}^\dagger_{i M_i}\big]^{\nu_{M_i}(A)}\,|0\rangle
\label{A-multiplets2-supplemental}\\
|n,i\rangle &\equiv& \big[\hat{f}^\dagger_{i 1}\big]^{\nu_{1}(n)}\!\!\!\!\!\!.\,.\,.\;
\big[\hat{f}^\dagger_{i M_i}\big]^{\nu_{M_i}(n)}\,|0\rangle\,.
\label{n-multiplets2-supplemental}
\eea
We call this Fock space ``embedding system'',
and expand the most general of its vectors as follows:
\be
\ket{\Phi_i} \equiv \sum_{A n} e^{i\frac{\pi}{2}N_n(N_n-1)}
\left[\phi_i\right]_{A n}\,U_{\text{PH}}\,
\ket{A,i}\ket{n,i}
\,,\label{pure-emb-supplemental}
\ee
where $N_n$ is the number of electrons in $\ket{n,i}$ and
$U_{\text{PH}}$ is the particle-hole (PH)
transformation satisfying the following identities,
\bea
U_{\text{PH}}^\dagger\,\hat{f}^\dagger_{ia}\,U_{\text{PH}}^\dagga &=& \hat{f}_{ia}\\
U_{\text{PH}}^\dagger\,\hat{f}_{ia}\,U_{\text{PH}}^\dagga &=& \hat{f}_{ia}^\dagger\\
U_{\text{PH}}^\dagger\,\hat{c}^\dagger_{i\alpha}\,U_{\text{PH}}^\dagga &=& \hat{c}^\dagger_{i\alpha}\\
U_{\text{PH}}^\dagger\,\hat{c}_{i\alpha}\,U_{\text{PH}}^\dagga &=& \hat{c}_{i\alpha}\,,
\eea
i.e., acting only on the $\hat{f}$ degrees of freedom.

Let us consider the embedding states such that
the matrix $\phi_i$ appearing in Eq.~\eqref{pure-emb-supplemental}
couples only states with $N_A=N_n$, i.e., that:
\be
\hat{N}_i^{\text{tot}}\,\ket{\Phi_i}=M_i\,\ket{\Phi_i}\,,
\label{n-blocks-supplemental}
\ee
where
\be
\hat{N}_i^{\text{tot}}\equiv 
\sum_a\hat{f}^\dagger_{i a}\hat{f}^\dagga_{i a} +
\sum_\alpha\hat{c}^\dagger_{i \alpha}\hat{c}^\dagga_{i \alpha}
\ee
is the total number operator in the embedding system $\mathcal{E}_{i}$,
and $M_i$ is the number of spin-orbitals
in the $R,i$ space.
By identifying the matrix $\phi_i$ of Eq.~\eqref{pure-emb-supplemental}
satsfying the properties defined above with the RISB amplitudes,
we have defined a one-to-one mapping between the space
of RISB amplitudes $\phi_i$ and the states $\ket{\Phi_i}$
of the embedding system.
As pointed out in Ref.~\onlinecite{Our-PRX-supplemental}, within this representation the RISB Lagrange
function [Eq.~\eqref{Lag-SB-supplemental}] can be rewritten as follows:
\bea
&&\Lag_{\text{SB}}[
\ket{\Phi},E^c;\,  \R,\R^\dagger,\lambda;\, \D,\D^\dagger, \lambda^{c};\,\Delta_p]=
-\lim_{\mathcal{T}\rightarrow 0}\frac{\mathcal{T}}{\mathcal{N}}\sum_{k}\sum_{m\in\mathbb{Z}}
\Tr\log\!
\left(\frac{1}{i (2m+1)\pi\mathcal{T}-\R\epsilon_{k}\R^\dagger-\lambda-\eta+\mu}\right)
e^{i(2m+1)\pi\mathcal{T} 0^+}
\nonumber\\&&\quad\quad
+\sum_i\left[\Av{\Phi_i}{\h_i^{\text{emb}}[\D_i,\D_i^\dagger;\lambda_i^c]}
+E^c_i\!\left(1-\langle \Phi_i | \Phi_i \rangle
\right)\right]
\nonumber\\&&\quad\quad
-\sum_i\left[
\sum_{ab}\left(
\left[\lambda_i\right]_{ab}+\left[\lambda^c_i\right]_{ab}\right)\left[\Delta_{pi}\right]_{ab}
+\sum_{c a\alpha}\left(
\left[\D_{i}\right]_{a\alpha}\left[\R_{i}\right]_{c\alpha}
\left[\Delta_{pi}(1-\Delta_{pi})\right]^{\frac{1}{2}}_{ca}
+\text{c.c.}\right)
\right]\,,
\label{Lag-SB-emb-supplemental}
\eea
where
\bea
&&\h_i^{\text{emb}}[\D_i,\lambda_i^c]\equiv
\h^{\text{loc}}_i[\{\hat{c}^\dagger_{i \alpha}\},\{\hat{c}^\dagga_{i \alpha}\}] +
\nonumber\\&&\;\,
\sum_{a\alpha} \left(
\left[\D_{i}\right]_{a\alpha}
\hat{c}^\dagger_{i \alpha}\hat{f}^\dagga_{i a}+\text{H.c.}\right)
+\sum_{ab} \left[\lambda^c_{i}\right]_{ab}
\hat{f}^\dagga_{i b}\hat{f}^\dagger_{i a}\label{h-emb-i-supplemental}
\eea
and $\ket{\Phi_i}$ is
an eigenstate of $\hat{N}_i^{\text{tot}}$ with eigenvalue $M_i$,
see Eq.~\eqref{n-blocks-supplemental}.

\subsubsection{1.~~~Unitary transformations of $\phi_i$} 

For later convenience it is useful to express the action
of a unitary similarity transformation of $\phi_i$
\be
\phi_i\,\longrightarrow\,X^\dagger\,\phi_i\,X
\label{similarity-phi-supplemental}
\ee
in terms of the corresponding embedding state $\ket{\Phi_i}$.
A direct calculation shows that, if we assume that
\be
\left[X,\sum_{\alpha=1}^{M_i} F^\dagger_{i\alpha}F^\dagga_{i\alpha}\right]=0
\,,
\ee
applying Eq.~\eqref{similarity-phi-supplemental}
to $\phi_i$ amounts to apply the following unitary
operator to the corresponding embedding state:
\be
\ket{\Phi_i}\,\longrightarrow\,\mathcal{X}^\dagger\,\ket{\Phi_i}\,,
\ee
where
\be
\mathcal{X}^\dagger \equiv X^\dagger \,\otimes\,
U^\dagga_{\text{PH}}\, ^tX\, U^\dagger_{\text{PH}}
\label{emb-X-general-supplemental}
\ee
and ``$\otimes$'' indicates the tensor product between an operator
acting only onto the $\hat{c}$ degrees of freedom (left)
and an operator acting only onto the $\hat{f}$ degrees of freedom (right).

Let us now assume that $X$
is a single-particle unitary transformation represented as
\be
X=e^{i\sum_{\alpha\beta}\xi_{\alpha\beta}\,F^\dagger_{i\alpha}F^\dagga_{i\beta}}
\label{X-xi-supplemental}
\ee
and $x$ is its restriction within the corresponding single-particle space.
Under this assumption Eq.~\eqref{emb-X-general-supplemental} reduces to
\be
\mathcal{X}^\dagger =
e^{i\sum_a\xi_{aa}}\,
e^{-i\sum_{\alpha\beta}\xi_{\alpha\beta}\,
\left[
\hat{c}^\dagger_{i\alpha}\hat{c}^\dagga_{i\beta}+\hat{f}^\dagger_{i\alpha}\hat{f}^\dagga_{i\beta}
\right]}
\,,
\label{emb-X-sp-supplemental}
\ee
which is a single-particle unitary transformation
acting on the $\hat{c}$ and $\hat{f}$ ladder operators as follows:
\bea
\mathcal{X}^\dagger\,\hat{c}^\dagger_{i\alpha}\,\mathcal{X}
&\equiv& \sum_{\alpha'}x^\dagger_{\alpha'\alpha}\,\hat{c}^\dagger_{i\alpha'}
\label{xc1-supplemental}\\
\mathcal{X}^\dagger\,\hat{f}^\dagger_{ia}\,\mathcal{X}
&\equiv& \sum_{a'}x^\dagger_{a'a}\,\hat{f}^\dagger_{ia'}
\label{xc2-supplemental}
\,.
\eea

In summary, we have shown that applying a similarity single-particle
unitary transformation to $\phi_i$, see Eq.~\eqref{similarity-phi-supplemental},
is equivalent to apply the single-particle unitary operator
[Eq.~\eqref{emb-X-sp-supplemental}]
to the corresponding embedding state $\ket{\Phi_i}$, which satisfies
Eqs.~\eqref{xc1-supplemental} and \eqref{xc2-supplemental}.
Note that, unless $\xi$ is traceless, the vacuum state of the embedding
system acquires a phase under this transformation.

\subsubsection{2.~~~Change of basis}

For later convenience, it is useful to show how
$\h_i^{\text{emb}}$ transforms under under changes of basis.
It can be readily verified using Eqs.~\eqref{tb-Hloc-supplemental},
\eqref{tb3-supplemental} and \eqref{tb6-supplemental} that
\be
\h_i^{\text{emb}}\,\longrightarrow\,
\bar{L}^{\text{emb}\dagger}_{i}\,\h_i^{\text{emb}}\,\bar{L}_{i}^{\text{emb}}
\ee
where $\bar{L}^{\text{emb}}_{i}$ is a single-particle unitary transformation
defined as follows:
\bea
\bar{L}^{\text{emb}\dagger}_{i}\,\hat{c}^\dagger_{i\alpha}\,\bar{L}_{i}^{\text{emb}}
&\equiv& \sum_{\alpha'}[L^\dagger_{i}]_{\alpha'\alpha}\,\hat{c}^\dagger_{i\alpha}\\
\bar{L}^{\text{emb}\dagger}_{i}\,\hat{f}^\dagger_{i\alpha}\,\bar{L}_{i}^{\text{emb}}
&\equiv& \sum_{\alpha'}[L^\dagger_{i}]_{\alpha'\alpha}\,\hat{f}^\dagger_{i\alpha}
\,.
\label{change-basis-emb-supplemental}
\eea
In particular, this observation implies that the eigenvalues of
$\h_i^{\text{emb}}$ are invariant under changes of basis.

By using the equations of Sec.~IV~1    
it can be readily
realized that applying the similarity transformation of Eq.~\eqref{tb1-supplemental} to the
matrix $\phi_i$ is equivalent to transform the corresponding embedding
vector $\ket{\Phi_i}$ as follows:
\be
\ket{\Phi_i}\,\longrightarrow\,\bar{L}^{\text{emb}\dagger}_{i}\,\ket{\Phi_i}\,.
\ee
Consequently, 
\bea
\Av{\Phi_i}{\h_i^{\text{emb}}} &\rightarrow&
\Av{\bar{L}^{\text{emb}\dagger}_{i}\,\Phi_i}
{\bar{L}^{\text{emb}\dagger}_{i}\,\h_i^{\text{emb}}\,\bar{L}_{i}^{\text{emb}}}
\nonumber\\
&=& \Av{\Phi_i}{\h_i^{\text{emb}}}
\,,
\eea
i.e., $\Av{\Phi_i}{\h_i^{\text{emb}}}$ is invariant under
changes of basis.
Note that this is expected, as Eq.~\eqref{tb1-supplemental} was constructed
in order to keep the value assumed by $\Lag_{\text{SB}}$ invariant.

\subsubsection{3.~~~Imposing the symmetries on $\ket{\Phi_i}$} 

Using the equations of Sec.~IV~1 
it can be verified that from the symmetry conditions
[Eqs.~\eqref{gtb3-supplemental} and \eqref{gtb6-supplemental}] it follows that
\be
[\bar{\gamma}_{in}, \h_i^{\text{emb}}]=0
\quad\forall n=1,..,h_i\,,\label{sym-Hemb-supplemental}
\ee
where $h_i$ is the order of the group $\bar{G}_{Ri}$, and
the operators $\bar{\gamma}_{in}$ are defined as
\be
\bar{\gamma}_{in}
\equiv \bar{g}_{Rin} \,\otimes\,
U^\dagga_{\text{PH}}\, ^t\bar{g}_{Rin}^\dagger\, U^\dagger_{\text{PH}}
\quad\forall g\in\bar{G}_{Ri}\label{gamma_in-supplemental}
\,,
\ee
and constitute a representation of the symmetry group $\bar{G}_{Ri}$ in
the embedding Hilbert space.
Similarly, it can be verified that 
the symmetry condition [Eq.~\eqref{sym-phi-supplemental}] can be rephrased
in terms of the vectors $\ket{\Phi_i}$ as follows:
\be
\bar{\gamma}_{in}\,\ket{\Phi_i}=\ket{\Phi_i}
\quad\forall n=1,..,h_i\,.
\label{sym-ketphi-supplemental}
\ee
Note that using Eq.~\eqref{sym-ketphi-supplemental} we can readily construct the
projector $\mathcal{P}_i$ onto the subspace of symmetric embedding states.
For discrete groups, in particular,
the projector over the symmetric states can
be represented as follows:
\be
\mathcal{P}_i\equiv \frac{1}{h_i}\sum_{n=1}^{h_i}\,\bar{\gamma}_{in}\,.
\ee

Let us now apply the equations derived above to characterize the
groups of rotations, which are particularly relevant in practice.
We observe that if $\bar{G}_{Ri}$ is a group of rotations
then all of the elements $\bar{g}_{in}$, see Eq.~\eqref{sym-phi-supplemental},
can be represented as in Eq.~\eqref{emb-X-sp-supplemental}:
\be
\bar{g}_{in} =
e^{i\sum_{\alpha\beta}\left[\sum_{k=1}^3\theta^k_{in}J^k_{i\alpha\beta}\right]\,F^\dagger_{i\alpha}F^\dagga_{i\beta}}
\,,
\ee
where $J^k_i$ are the generators of the rotations in the corresponding
single-particle space. Since $J^k_i$ are traceless, using Eq.~\eqref{emb-X-sp-supplemental}
we deduce that the corresponding representative $\bar{\gamma}_{in}$
acting on the embedding space can be represented as follows:
\be
\bar{\gamma}_{in} =
e^{i\sum_{\alpha\beta}\left[\sum_{k=1}^3\theta^k_{in}J^k_{i\alpha\beta}\right]\,
\left[
\hat{c}^\dagger_{i\alpha}\hat{c}^\dagga_{i\beta}+\hat{f}^\dagger_{i\alpha}\hat{f}^\dagga_{i\beta}
\right]}\,,
\ee
that is a rotation acting with the same Lie parameters $\theta^k_{in}$
both on the $\hat{c}$ and on the $\hat{f}$ degrees of freedom.

It is also interesting to observe that Eq.~\eqref{n-blocks-supplemental} can be deduced
as we did for the groups of rotations from the condition:
\be
\left[\phi_i,e^{\sum_{\alpha=1}^{M_i} F^\dagger_{i\alpha}F^\dagga_{i\alpha}\,\xi}\right]
=0\quad\forall\,\xi\,,
\label{N-cons-phi-supplemental}
\ee
which amounts to enforce the assumption that $\phi_i$ can couple only states
with the same number of electrons.
In fact, Eq.~\eqref{emb-X-sp-supplemental} enables us to represent Eq.~\eqref{N-cons-phi-supplemental} as
follows:
\be
e^{i\sum_a\xi\,M_i}\,
e^{-i\sum_{\alpha\beta}\xi\,\left[\hat{c}^\dagger_{i\alpha}\hat{c}^\dagga_{i\alpha}+\hat{f}^\dagger_{i\alpha}\hat{f}^\dagga_{i\alpha}\right]}
\,\ket{\Phi_i} = \ket{\Phi_i}\quad\forall\,\xi\,,
\ee
which is equivalent to Eq.~\eqref{n-blocks-supplemental}.

As we have shown above, the lowest-energy eigenspace of
$\h_i^{\text{emb}}$ is the basis of a 
representation of the $(R,i)$ point group of the system,
see Eq.~\eqref{gamma_in-supplemental}, which is presumably irreducible.
If the so obtained ground state is such that Eq.~\eqref{sym-ketphi-supplemental} is 
automatically verified, then it is not necessary to restrict the
search of the ground state of $\h_i^{\text{emb}}$ to the
subspace of symmetric states.
Indeed, in several cases we found convenient not to impose 
the symmetry conditions [Eq.~\eqref{sym-ketphi-supplemental}]
(or to impose them only for a subgroup of $\bar{G}_{Ri}$).
The reason is that, even though applying to $\h_i^{\text{emb}}$ the projector 
over the symmetric states effectively reduces
the dimensionality of the problem, in some case this operation compromises
considerably the sparsity of its representation.
In general, the most convenient option depends on the specific
system considered.
This technical detail will be discussed further
in Sec.~V~A.  

\section{V.~~~Solution of RISB Lagrange equations}

For later convenience we define
the projectors $\Pi_i$ over the single-particle $(R,i)$ local subspaces.
The symbol $f$ will indicate the Fermi function.

\subsection{A.~~~Variational setup} 

In order to take into account the symmetry conditions,
see 
Eqs.~\eqref{sym-phi-supplemental}-\eqref{sym-sp-2-supplemental}, and
the fact that $\Delta_{pi}$, $\lambda^c_i$ and $\lambda_i$ are
Hermitian matrices, we introduce the following parametrizations:
\bea
\Delta_{pi}&=&\sum_s d^p_{is}\,^th_{is}
\\
\lambda^c_i&=&\sum_s l_{is}^c\,h_{is}
\label{lambdac-lc-supplemental}
\\
\lambda_i&=&\sum_s l_{is}\,h_{is}
\label{lambda-l-supplemental}
\\
\R_i&=&\sum_sr_{is}\,h_{is}
\label{R-r-supplemental}
\eea
where the set of matrices $h_{is}$ is an orthonormal
basis of the space of Hermitian matrices with
dimension $M_i$ satisfying the symmetry conditions:
\be
[g_{Rin},h_{is}]=0
\quad\forall\bar{g}_{Rin}\in \bar{G}_{Ri}
\,,
\ee
and $d^p_{is}$, $l_{is}^c$ and $l_{is}$ are real numbers, while
$r_{is}$ are complex numbers.
The above-mentioned orthonormality is defined
with respect to the standard scalar product
$(A,B)\equiv\Tr\!\left[A^\dagger B\right]$.
Note that from the definitions above it follows that
\bea
\sum_{ab}\left(
\left[\lambda_i\right]_{ab}+\left[\lambda^c_i\right]_{ab}\right)\left[\Delta_{pi}\right]_{ab}&=&\sum_s(l_{is}+l^c_{is})\,d^p_{is} 
\equiv (l_i+l_i^c,d^p_{i})\,.
\label{lplc-d-supplemental}
\eea

As discussed in the previous section, the subspace $\mathcal{V}^{\text{E}}_i$
of symmetric embedding states $\ket{\Phi_i}$ is identified by
Eqs.~\eqref{n-blocks-supplemental} and \eqref{sym-ketphi-supplemental}.
Let us assume that we have calculated for each $i$ a basis
of $\mathcal{V}^{\text{E}}_i$:
\be
\mathcal{B}^{\text{E}}_i\equiv\{\ket{\Phi_{iS}}\,|\,S=1,...,D^{\text{E}}_i\}\,,
\ee
where $D^{\text{E}}_i$ is the dimension of $\mathcal{V}^{\text{E}}_i$.
Within these definitions, any symmetric embedding state can be expanded as
follows:
\be
\ket{\Phi_i} = \sum_{S=1}^{D^{\text{E}}_i} c_{iS}\, \ket{\Phi_{iS}}
\quad\forall\,\ket{\Phi_i}\in\mathcal{V}^{\text{E}}_i
\,,
\ee
where $c_{iS}$ are complex numbers.

In order to take
into account the symmetry conditions of $\ket{\Phi_i}$
it is sufficient to pre-calculate the following objects:
\bea
U^i_{SS'}&\equiv&\langle \Phi_{iS} |\,
\h^{\text{loc}}_i[\{\hat{c}^\dagger_{i \alpha}\},\{\hat{c}^\dagga_{i \alpha}\}]
\,| \Phi_{iS'} \rangle 
\label{t1-supplemental}\\
N^{i ab}_{SS'}&\equiv&\langle \Phi_{iS} |\,
\hat{f}^\dagga_{i b}\hat{f}^\dagger_{i a}
\,| \Phi_{iS'} \rangle 
\label{t2-supplemental}\\
M^{i a\alpha}_{SS'}&\equiv&\langle \Phi_{iS} |\,
\hat{c}^\dagger_{i \alpha}\hat{f}^\dagga_{i a}
\,| \Phi_{iS'} \rangle
\label{t3-supplemental}
\,,
\eea
which are the representations in the basis $\mathcal{B}^{\text{E}}_i$
of the ``components'' of $\h_i^{\text{emb}}$ projected
within the subspaces $\mathcal{V}^{\text{E}}_i$ of symmetric states.
In fact, using these definitions, we can express the matrix elements
of $\h_i^{\text{emb}}$ as follows:
\bea
\langle \Phi_{iS} |\,\h_i^{\text{emb}}\,| \Phi_{iS'}\rangle
&=&\sum_{a\alpha} \left[\D_{i}\right]_{a\alpha}\,M^{ia\alpha}_{SS'}+
\sum_{ab} \left[\lambda^c_{i}\right]_{ab}\,N^{iab}_{SS'}
+ U^i_{SS'}
\,.
\eea

Note that the representations [Eqs.~\eqref{t2-supplemental} and \eqref{t3-supplemental}] are
very sparse if $\mathcal{B}^{\text{E}}_i$ is made of Fock states.
It is for this reason that, as anticipated at
the end of Sec.~IV~3,   
in several cases it is convenient
not to impose all of the symmetry conditions of $\ket{\Phi_i}$ in
order to work in a Fock basis ---
even though doing so increases the dimension $D^{\text{E}}_i$ of the problem.

From now on we will define ``variational setup''
the set of matrices $h_{is}$, see
Eqs.~\eqref{lambdac-lc-supplemental}-\eqref{R-r-supplemental}, and the objects
represented in Eqs.~\eqref{t1-supplemental}-\eqref{t3-supplemental}.
In our current implementation the variational setup is pre-calculated
and stored on disk before to solve numerically
the RISB Lagrange equations.

We point out that if the RISB method is applied in combination
with LDA (LDA+RISB) it is necessary to store separately the representations
of the quadratic components of $\h^{\text{loc}}_i$ (crystal fields) and the quartic
part (interaction), as the crystal fields change at each charge iteration.

\subsection{B.~~~Gauge-invariant Lagrange Equations} 

It can be readily shown that the saddle-point conditions of $\Lag_{\text{SB}}$,
see Eq.~\eqref{Lag-SB-supplemental},
with respect to  all of its arguments provides the following system
of Lagrange equations:
\bea
&&\frac{1}{\mathcal{N}}\left[
\sum_k \Pi_i f\!\left(\R\epsilon_{k}\R^\dagger+\lambda\right)\Pi_i\right]_{ba}
 = \left[\Delta_{pi}\right]_{ab}\label{l2-supplemental}\\
&&\frac{1}{\mathcal{N}}\left[\frac{1}{\R_i}
\sum_k \Pi_i\,\R\epsilon_k\R^\dagger\,
f\!\left(\R\epsilon_k\R^\dagger+\lambda\right)\Pi_i\right]_{\alpha a}
=\sum_c\left[\D_{i}\right]_{c\alpha}
\left[\Delta_{ip}\left(1-\Delta_{ip}\right)\right]^{\frac{1}{2}}_{ac}
\label{l3-supplemental}\\
&& \sum_{cb\alpha}\frac{\partial}{\partial d^p_{is}}\left[\Delta_{pi}\left(1-\Delta_{pi}\right)\right]^{\frac{1}{2}}_{cb} [\D_i]_{b\alpha}[\R_i]_{c\alpha}+\text{c.c.}
+\left[l+l^c\right]_{is}=0
\label{l4-supplemental}\\
&&\h_i^{\text{emb}}[\D_i,\lambda_i^c]\,\ket{\Phi_i}
= E^c_i\,\ket{\Phi_i}
\label{l5-supplemental}\\
&&\left[\mathcal{F}^{(1)}_i\right]_{\alpha a}\equiv
\Av{\Phi_i}{\hat{c}^\dagger_{i \alpha}\hat{f}^\dagga_{i a}}
-\sum_c\left[\Delta_{ip}\left(1-\Delta_{ip}\right)\right]^{\frac{1}{2}}_{ca} [\R_i]_{c\alpha}
=0\label{l6-supplemental}\\
&&\left[\mathcal{F}^{(2)}_i\right]_{ab}\equiv
\Av{\Phi_i}{\hat{f}^\dagga_{i b}\hat{f}^\dagger_{i a}}
- \left[\Delta_{pi}\right]_{ab}
=0 \label{l7-supplemental}\,.
\eea
Note that the projectors $\Pi_i$ appear in Eq.~\eqref{l3-supplemental}
because derivatives are taken with respect to the matrix elements
of the block matrices $\eta$, $\lambda_i$ and $\R_i$,
and that Eq.~\eqref{lplc-d-supplemental} has been used to obtain
Eq.~\eqref{l4-supplemental}.
The partial derivative with respect to $d^p_{is}$ of
$\left[\Delta_{pi}\left(1-\Delta_{pi}\right)\right]^{\frac{1}{2}}$ can be
calculated semi-analytically in several ways, see, e.g.,
Ref.~\onlinecite{positive-definite-matrices-supplemental}.

A possible way to compute the solution is the
following~\cite{Gmethod-supplemental}.
(I) Given a set of coefficients $r_{is}$ and $l_{is}$, we determine the
corresponding matrices $\R$ and $\lambda$ using
Eqs.~\eqref{lambda-l-supplemental} and \eqref{R-r-supplemental}, and calculate
$\Delta_{pi}$ using Eq.~\eqref{l2-supplemental}.
(II) We calculate $\D_i$ by inverting Eq.~\eqref{l3-supplemental}.
(III) We calculate the coefficients $l^c_{is}$ using Eq.~\eqref{l4-supplemental}
and the corresponding matrix $\lambda^c_i$ using Eq.~\eqref{lambdac-lc-supplemental}.
(IV) We construct the embedding Hamiltonian $\h_i^{\text{emb}}$
and compute its ground state $\ket{\Phi_i}$, see Eq.~\eqref{l5-supplemental},
within the subspace identified by
Eqs.~\eqref{n-blocks-supplemental} and \eqref{sym-ketphi-supplemental}.
(V) We determine the left members of Eqs.~\eqref{l6-supplemental} and \eqref{l7-supplemental}.
The equations \eqref{l6-supplemental} and \eqref{l7-supplemental} are satisfied
if and only if the coefficients $r_{is}$ and $l_{is}$ proposed
at the first of the steps above
identify a solution of the RISB Lagrange function.

In conclusion, we have formulated the solution of the RISB equations
as a root problem for a function of
$\left(r_{is},l_{is}\right)$,
which can be formally represented as follows:
\be
\mathcal{F}(r,l)\equiv
\left(\mathcal{F}_1(r,l),...,\mathcal{F}_{n_c}(r,l)\right)=0
\label{roothubb-supplemental}
\ee
where $n_c$ is the number of atoms within the unit cell and
\be
\mathcal{F}_i(r,l)\equiv
\left(
\mathcal{F}_i^{(1)}(r,l),\mathcal{F}^{(2)}_i(r,l)
\right)
=0\quad\forall i\,.
\ee

Eq.~\eqref{roothubb-supplemental} can be
solved numerically, e.g., using the quasi-Newton method.
We remark that, as pointed out in Ref.~\onlinecite{Our-PRX-supplemental}, each
component $\mathcal{F}_i$ of the the vector-function $\mathcal{F}$
can be evaluated independently
through the numerical steps outlined above.

\subsection{C.~~~Restarting calculations in the presence of a symmetry-breaking perturbation} 

Let us consider a generic RISB
Hamiltonian $\hat{H}$ defined by the parameters
$\epsilon_k$ and $H^{\text{loc}}_i$, see Eq.~\eqref{hu-supplemental},
and assume that it is
invariant with respect to the point groups $G_i$ (a
point group for each atom $i$ within the unit cell).

In Sec.~III 
we have shown that
the symmetry conditions to be satisfied by
the RISB variational parameters
depend on the representations $\bar{G}_i$ of $G_i$, see
Eq.~\eqref{sp-g-supplemental}.
Using these representations,
in Sec.~V~A  
we have introduced:
(i) the set of matrices $h_{is}$, see
Eqs.~\eqref{lambdac-lc-supplemental}-\eqref{R-r-supplemental}, and
(ii) the tensors $U$, $M$ and $N$ represented in Eqs.~\eqref{t1-supplemental}-\eqref{t3-supplemental}.
These objects constitute the so called variational setup,
and encode all of the symmetry conditions to be
enforced on the RISB variational parameters.

In summary, 
the input parameters defining the RISB Lagrange equations of $\hat{H}$,
see Eqs.~\eqref{l2-supplemental}-\eqref{l7-supplemental}, are the following:
(1) the parameters of the Hamiltonian $\epsilon_k$ and $H^{\text{loc}}_i$, and
(2) the above mentioned variational setup.
For later convenience, let us make these dependencies of
Eq.~\eqref{roothubb-supplemental} explicit as follows:
\be
\mathcal{F}^{\epsilon_k}_{h_{is};U^i,N^i,M^i}(r,l) = 0\,.
\label{roothubb-pars-supplemental}
\ee
Note that $H^{\text{loc}}_i$ does not appear explicitly
in Eq.~\eqref{roothubb-pars-supplemental},
as all we need in practice is its projection within the space
of symmetric embedding states, which is encoded within the 
variational setup tensor $U^i$.

As anticipated at the end of Sec.~III~C,   
the fact that the gauge-invariant Lagrange function is explicitly covariant
under changes of basis makes it easier to solve systems
constituted by a main term with high symmetry and a
smaller perturbation breaking part of it.
In this section we derive a convenient method to solve this problem.

We consider a Hubbard Hamiltonian represented as
\be
\hat{H}=\hat{H}^0+\delta\hat{H}\,,
\label{split-hamgen-supplemental}
\ee
where $\hat{H}^0$ is invariant with respect to the point groups $G^0_i$,
while $\delta\hat{H}$ is a ``small''
perturbation invariant only with respect to the subgroups $G_i\subset G^0_i$.
Consistently with Eq.~\eqref{split-hamgen-supplemental}, the parameters defining 
Eq.~\eqref{hu-supplemental} are represented as
\bea
\epsilon_k &=& \epsilon^0_k + \delta\epsilon_k\\
H^{\text{loc}}_i &=& H^{0\,\text{loc}}_i + \delta H^{\text{loc}}_i\,.
\eea

Let us represent schematically the ``unperturbed'' Lagrange
equations as follows:
\be
\mathcal{F}^{\epsilon^0_k}_{h^0_{is};U^{0i},N^{0i},M^{0i}}(r^0,l^0) = 0\,.
\label{roothubb-pars0-supplemental}
\ee
Since $\hat{H}^0$ has (by assumption) more symmetries than the full Hamiltonian,
the Lagrange equations represented
by Eq.~\eqref{roothubb-pars0-supplemental} are simpler to solve.
The reasons are the following.
(1) The number of symmetric matrices $h^0_{is}$ --- which is equal
to the dimension of $r^0$ and $l^0$ --- is smaller.
This reduces the number of evaluations of [Eq.~\eqref{roothubb-pars0-supplemental}] necessary
to solve the root problem.
(2) The dimension of the tensors $U^{0i}$, $N^{0i}$ and $M^{0i}$ is smaller.
This reduces the computational coast of calculating the ground state
of $\h_i^{\text{emb}}$, which is generally the most time consuming operation
necessary in order to evaluate the function [Eq.~\eqref{roothubb-pars0-supplemental}].

It is important to observe that, thanks to the covariance of the
RISB Lagrange equations, the space generated by $h^0_{is}$ is a
well defined \emph{subspace}
of the space generated by $h_{is}$, see  Eq.~\eqref{roothubb-pars-supplemental}.
Consequently, Eq.~\eqref{roothubb-pars0-supplemental} can be viewed as an
approximation to the restriction of Eq.~\eqref{roothubb-pars-supplemental}
within a subspace of $(r,l)$, where
\be
\delta\F\equiv
\mathcal{F}^{\epsilon^k}_{h_{is};U^{i},N^{i},M^{i}}-
\mathcal{F}^{\epsilon^0_k}_{h^0_{is};U^{0i},N^{0i},M^{0i}}
\ee
is presumably small if $\delta\h$ is small.
Thanks to this observation,
we can use the solution of the unperturbed problem [Eq.~\eqref{roothubb-pars0-supplemental}]
as a starting point for the quasi-Newton solver, thus speeding up
the solution of
the root problem in the presence of $\delta\h$, see
Eq.~\eqref{roothubb-pars-supplemental}.

\section{VI.~~~Other numerical advantages of the gauge-invariant formulation}

In this section we discuss a few more differences between the numerical solution
of the gauge invariant RISB Lagrange functions [Eq.~\eqref{Lag-SB-supplemental}]
the Lagrange function of Ref.~\onlinecite{Our-PRX-supplemental}, which amounts to
fix the gauge in which $\Delta_p$ is diagonal (natural basis).

In order to illustrate these differences,
let us write explicitly the saddle point conditions of
the natural-basis Lagrange function of Ref.~\onlinecite{Our-PRX-supplemental}:
\bea
&&\frac{1}{\mathcal{N}}\left[
\sum_k \Pi_i f\!\left(\R\epsilon_{k}\R^\dagger+\lambda+\eta\right)\Pi_i\right]_{ba}
 = 0 \;\,\forall a\neq b\label{leta-supplemental}\\
&&\frac{1}{\mathcal{N}}\left[
\sum_k \Pi_i f\!\left(\R\epsilon_{k}\R^\dagger+\lambda+\eta\right)\Pi_i\right]_{ba}
 = \left[n^0_{i}\right]_{ab}\label{l2-ga-supplemental}\\
&&\frac{1}{\mathcal{N}}\left[\frac{1}{\R_i}
\sum_k \Pi_i\,\R\epsilon_k\R^\dagger\,
f\!\left(\R\epsilon_k\R^\dagger+\lambda+\eta\right)\Pi_i\right]_{\alpha a}
=\left[\D_{i}\right]_{a\alpha}
\sqrt{\left[n^0_{i}\right]_{aa}\left(1-\left[n^0_{i}\right]_{aa}\right)}
\label{l3-ga-supplemental}\\
&& \frac{\left[n^0_{i}\right]_{aa}\!-\!\frac{1}{2}}{\sqrt{\left[n^0_{i}\right]_{aa}\left(1-\left[n^0_{i}\right]_{aa}\right)}}
\left[\sum_\alpha\left[\D_{i}\right]_{a\alpha}\left[\R_{i}\right]_{a\alpha}\!+\!\text{c.c.}\right]\delta_{ab}
-\left[\lambda_i\!+\!\lambda^c_i\right]_{ab}=0\label{l4-ga-supplemental}\\
&&\h_i^{\text{emb}}[\D_i,\lambda_i^c]\,\ket{\Phi_i}
= E^c_i\,\ket{\Phi_i}
\label{l5-ga-supplemental}\\
&&\left[\mathcal{F}^{(1)}_i\right]_{\alpha a}\equiv
\Av{\Phi_i}{\hat{c}^\dagger_{i \alpha}\hat{f}^\dagga_{i a}}
-\left[\R_{i}\right]_{\alpha a}
\sqrt{\left[n^0_{i}\right]_{aa}\left(1-\left[n^0_{i}\right]_{aa}\right)}
=0\label{l6-ga-supplemental}\\
&&\left[\mathcal{F}^{(2)}_i\right]_{ab}\equiv
\Av{\Phi_i}{\hat{f}^\dagga_{i b}\hat{f}^\dagger_{i a}}
 - \left[n^0_{i}\right]_{ab}
=0 \label{l7-ga-supplemental}\,.
\eea

Note that also in the natural-basis gauge-fixing formulation
of the RISB method the numerical problem amounts to solve a root problem
represented as in Eq.~\eqref{roothubb-supplemental}.
However, as we are going to show, the gauge-invariant formulation
presents several numerical advantages.

The most important advantage of the gauge-invariant formulation,
which was already mentioned in the main text,
is that, while the number of independent variables defining $\R$ and $\lambda$,
--- which are the arguments of the root problem
[Eq.~\eqref{roothubb-supplemental}] to be solved --- is identical in the two approaches,
within the gauge-invariant formulation there exists a manifold of
physically equivalent solutions, which are mapped one onto the other
by gauge transformations, see Eq.~\eqref{gauge-R-supplemental}.
The above-mentioned multiplicity of solutions effectively reduces the dimension
of the root problem, and turns out to considerably speed up convergence
by reducing considerably the number of evaluations of $\mathcal{F}_i$ necessary to
solve it.

Another important advantage of the gauge-invariant formulation is that
it is not necessary to solve numerically Eq.~\eqref{leta-supplemental},
which consists in applying the natural-basis gauge fixing.
Note that when the method is applied within the framework of LDA+RISB 
this operation can be very time consuming.
In fact, since the single-particle Hilbert
space contains also the uncorrelated orbitals,
the matrix $\epsilon_k$ has generally a relatively large dimension.

\section{VII.~~~Supplemental details about electronic structure of UO$_2$}

\subsection{A.~~~Parametrization Slater-Condon Interaction}

As discussed in the main text, in our calculations of UO$_2$
we employed the following parameters
for the Slater-Condon local interaction: $U=10\,eV$, $J=0.6\,eV$.
Here we clarify the how these values were used to
parameterize the Slater integrals.

As discussed in Ref.~\onlinecite{LDA+U-supplemental}, for $f$-electrons
the Coulomb $U$ and Hund's $J$ parameters are related with
the Slater integrals as follows: $U=F^0$,
and $J=(286F^2+195F^4+250F^6)/6453$.
Following Ref.~\onlinecite{LDA+U-supplemental}, in our work we assumed the following
ratios between the Slater integrals, which are known to
hold with good accuracy for $f$-systems: 
$F^4/F^2\simeq 0.668$, $F^6/F^2\simeq 0.494$.
These conditions enable us to express all of the Slater integrals in terms of
only $U$ and $J$.

\subsection{B.~~~Procedure of Averaging over the Crystal Field Splittings (CFS)}

In our DFT+RISB calculations we have fully taken into account both spin
orbit and CFS.
However, as discussed in the main text, in order to evaluate the importance of the CFS
we have compared our results with those obtained by \emph{``averaging over the CFS''}.
For completeness, here we describe in detail the averaging procedure.


As discussed above, our approach to solve the RISB mean field equations,
see Eqs.~\eqref{l2-supplemental}-\eqref{l7-supplemental},
consists in a root problem in the parameters $(\R,\lambda)$, which
encode the RISB self-energy as follows~\cite{Our-PRX-supplemental}:
\be
\Sigma(\omega)=- \omega\,\frac{I-\R^\dagger\R}{\R^\dagger\R}
+\frac{1}{\R}\lambda\frac{1}{\R^\dagger}\,.
\ee
In particular, this procedure requires to solve
recursively the ``embedding Hamiltonian'' [Eq.~\eqref{h-emb-i-supplemental}],
which is an impurity model where the bath
has only the same dimension of the impurity.

The details of the above-mentioned procedure of
``averaging over the CFS'' is defined as follows.
\begin{itemize}
	
	\item[1] The above-mentioned root problem is solved by restricting the search of
	parameters $(\R,\lambda)$ assuming that $[\R,\mathbf{J}]=[\lambda,\mathbf{J}]=0$,
	where $\mathbf{J}=\mathbf{L}+\mathbf{S}$ is the total angular momentum.
	Thus, both of the averaged matrices are diagonal and have only 2 independent components
	labeled by the corresponding eigenvalues of $J^2$, i.e., $5/2$ and $7/2$.
	
	\item[2] Similarly, the left members of Eqs.~\eqref{l2-supplemental} and \eqref{l3-supplemental} are fitted
	(at each iteration)
	to an isotropic form, i.e., to a form
	diagonal with only 2 independent components labeled by $5/2$ and $7/2$
	(which is equivalent to assume that the environment of the impurity of the
	embedding Hamiltonian is isotropic).
	
	\item[3] Also the ``on-site energies'', i.e., the quadratic part of the U-$5f$
	local Hamiltonian (which is incorporated in the impurity component of the
	embedding Hamiltonian [Eq.~\eqref{h-emb-i-supplemental}] and is determined by LDA)
	is fitted to an isotropic form at each iteration. Note that this amounts to
	neglect the splittings of the on-site energies due to the crystal fields. 
	
\end{itemize}

Physically, the averaging procedure described above amounts to assume that
the U-$5f$ degrees of freedom of each U atom
can be approximately treated as if their environment was isotropic
--- which would be the case if the CFS were negligible.
As discussed in the main text, the comparison between the full calculations and those
obtained by ``averaging over the CFS'' enabled us to clarify that
the taking CFS into account is essential in UO$_2$,
as the averaging procedure results into a
description of the electronic structure which is unphysical in many
respects --- such as the pattern of orbital differentiation of
this material.

As discussed in the main text, in
order to investigate the physical origin of the importance of the CFS,
the calculations were repeated
also by performing the averaging procedure only over the impurity levels of the
impurity Hamiltonian, see the point (c) above.
The fact that performing the averaging procedure only on the on-site energies
did not affect sensibly the result of our calculations
enabled us to deduce that the underlying reason why the CFS are important in UO$_2$
concerns the hybridization mechanism between the U-$5f$ and O-$2p$ degrees of freedom,
and not the consequent splittings of the on-site impurity energy-levels, which are,
in fact, very small in this material.

\subsection{C.~~~Calculation Orbital Occupations of Table~I of main text}

Here we point out that the physical occupations reported in Table~I of
the main text were calculated directly from the RISB wavefunction
[Eq.~\eqref{varsb-supplemental}] as follows.

Let us consider the density-matrix operators:
\be
\hat{\rho}_{\alpha\beta}\equiv\cc_{Ri\alpha}\ca_{Ri\beta}
\equiv \sum_{AB} [F^\dagger_{i \alpha}F^\dagga_{i \beta}]_{AB}\,\ket{A,Ri}\bra{B,Ri}\,,
\ee
where the matrices $F_{i\alpha}$ and the operators $\ket{A,Ri}\bra{B,Ri}$
were defined in the main text.
Within the operatorial RISB representation derived in this work,
similarly to Eq.~\eqref{hloc_bosons-supplemental},
the operators $\hat{\rho}_{\alpha\beta}$
can be represented as follows:
\be
\hat{\rho}_{\alpha\beta}\equiv\cc_{Ri\alpha}\ca_{Ri\beta}\;\longrightarrow\;
\sum_{AB}\,
[F^\dagger_{i \alpha}F^\dagga_{i \beta}]_{AB}\sum_n\Phi^\dagger_{RiAn}\Phi^\dagga_{RiBn}\,.
\ee

The expectation value of the above operators with respect to the mean-field
wavefunction [Eq.~\eqref{varsb-supplemental}] is given by:
\be
\Av{\Psi_{\text{SB}}}{\sum_{AB}\,
	[F^\dagger_{i \alpha}F^\dagga_{i \beta}]_{AB}\sum_n\,\Phi^\dagger_{RiAn}\Phi^\dagga_{RiBn}}
= \Tr\big[
\phi^\dagga_{i}\phi^\dagger_{i}\,F^\dagger_{i\alpha}F^\dagga_{i\beta}
\big]\,,
\label{occupations}
\ee
which is entirely expressed in terms of the SB amplitudes.

Note that, since $\phi^\dagger_{i}$ and $\phi_{i}$ do \emph{not} commute,
\be
\Tr\big[
\phi^\dagga_{i}\phi^\dagger_{i}\,F^\dagger_{i\alpha}F^\dagga_{i\beta}
\big]\,\neq\,
\Tr\big[
\phi^\dagger_{i}\phi^\dagga_{i}\,F^\dagger_{i\alpha}F^\dagga_{i\beta}
\big]\,.
\ee
Consequently, the physical occupations represented in
Eq.~\eqref{occupations} are not directly related with the so-called
quasi-particle occupations appearing in Eq.~\eqref{avC2-supplemental}.

\subsection{D.~~~Energetics UO$_2$}

\begin{figure}
	\begin{center}
		\includegraphics[width=8.4cm]{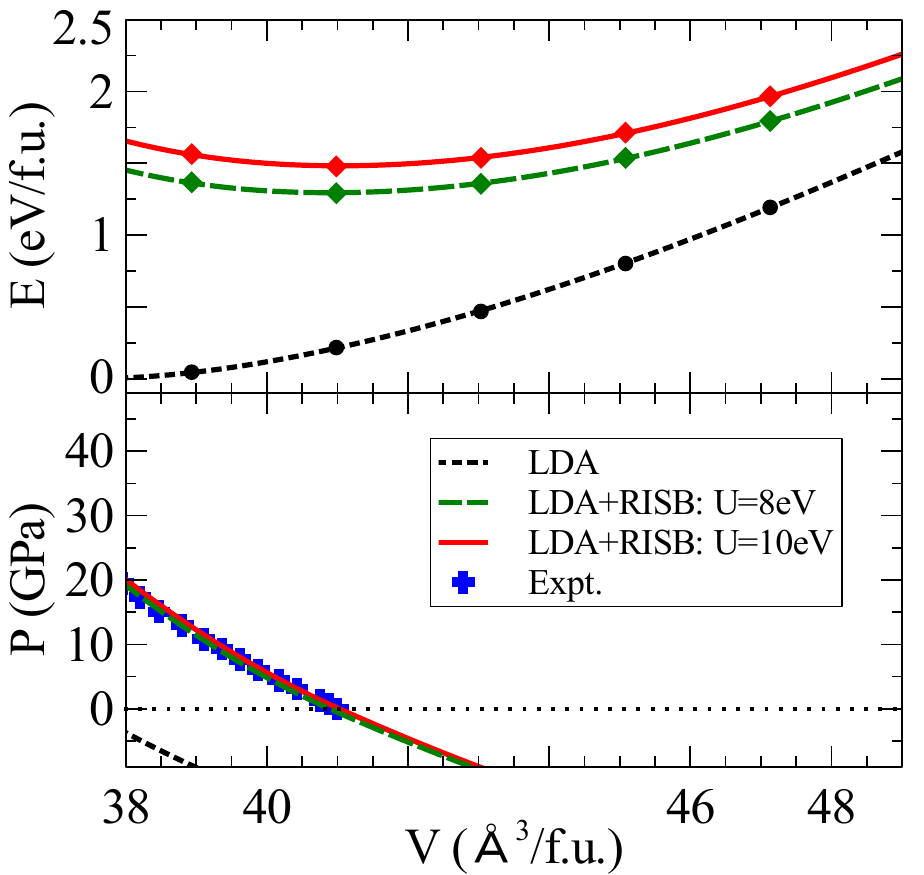}
		\caption{(Color online) 
			Zero temperature LDA and LDA+RISB total energies (upper panel)
			and corresponding pressure-volume phase diagrams
			compared with the room-temperature experiments of Ref.~\onlinecite{UO2-exp-supplemental}
			(lower panel).
		}
		\label{figure1}
	\end{center}
\end{figure}

In the upper panel of Fig.~\ref{figure1} are shown the LDA and LDA+RISB
total energies $E(V)$ obtained at zero temperature for $U=8\,eV$ and 
$U=10\,eV$.
The corresponding pressure (P-V) curves, obtained from $P(V)=-dE/dV$,
are shown in the lower panel in comparison with
the experimental data of Ref.~\onlinecite{UO2-exp-supplemental} (which were obtained
at room temperature).
As anticipated in the main text, we observe that
the P-V curve (and, in particular, the equilibrium volume) is essentially
identical for $U=8\,eV$, as changing $U$ results in an energy shift
that is essentially volume independent.
The agreement with the experiment is remarkably good with
both of the values of $U$ considered.

\subsection{E.~~~Full matrix quasi-particle weights UO$_2$}

For completeness, below we report the complete representation
of the matrix of quasi-particle weights
$Z=\R^\dagger\R$ of the U-$5f$ electrons in the basis [Eq.~16] of 
the main text:
\[
Z_{\Gamma_8} = 
\begin{bmatrix}
\ket{\Gamma_8^{(1)},5/2,+} & \ket{\Gamma_8^{(2)},7/2,-} & \ket{\Gamma_8^{(1)},5/2,-} & \ket{\Gamma_8^{(2)},7/2,+} & \ket{\Gamma_8^{(2)},5/2,+} & \ket{\Gamma_8^{(1)},7/2,-} & \ket{\Gamma_8^{(2)},5/2,-} & \ket{\Gamma_8^{(1)},7/2,+} \\
\hline
0.1079 & 0.2952 & 0 & 0 & 0 & 0 & 0 & 0 \\
0.2952 & 0.8073 & 0 & 0 & 0 & 0 & 0 & 0 \\
0      & 0      & 0.1079 & 0.2952 & 0 & 0 & 0 & 0 \\
0      & 0      & 0.2952 & 0.8073 & 0 & 0 & 0 & 0 \\
0      & 0      & 0      & 0      & 0.1079 & 0.2952 & 0 & 0 \\
0      & 0      & 0      & 0      & 0.2952 & 0.8073 & 0 & 0 \\
0      & 0      & 0      & 0      & 0      & 0 & 0.1079 & 0.2952 \\
0      & 0      & 0      & 0      & 0      & 0 & 0.2952 & 0.8073 
\end{bmatrix}
\]

\[
Z_{\Gamma_7} = 
\begin{bmatrix}
\ket{\Gamma_7,5/2,+} & \ket{\Gamma_7,7/2,-} & \ket{\Gamma_7,5/2,-} & \ket{\Gamma_7,7/2,+} \\
\hline
0.9244 & 0.0125 & 0 & 0 \\
0.0125 & 0.9489 & 0 & 0 \\
0      & 0      & 0.9244 & 0.0125 \\
0      & 0      & 0.0125 & 0.9489
\end{bmatrix}
\;\;\;
Z_{\Gamma_6} =
\begin{bmatrix}
\ket{\Gamma_6,7/2,+} & \ket{\Gamma_6,7/2,-} \\
\hline
0.9515 & 0 \\
0      & 0.9515 
\end{bmatrix}
\]
Because of the Schur lemma, the states belonging to inequivalent representations
are not coupled by the self energy (and, consequently, by $Z$).
Note that, as discussed in the main text,
the off-diagonal matrix elements of $Z$ coupling $5/2$ and $7/2$ states
are not negligible.

\subsection{F.~~~Single-particle density matrix UO$_2$}

Below we report the complete representation
of the single-particle density matrix
$\rho_{\alpha\beta}=\langle \cc_\alpha\ca_\beta \rangle$
of the U-$5f$ electrons in the basis [Eq.~16] of 
the main text:
\[
\rho_{\Gamma_8} =
\begin{bmatrix}
\ket{\Gamma_8^{(1)},5/2,+} \!&\! \ket{\Gamma_8^{(2)},7/2,-} \!&\! \ket{\Gamma_8^{(1)},5/2,-} \!&\! \ket{\Gamma_8^{(2)},7/2,+} \!&\! \ket{\Gamma_8^{(2)},5/2,+} \!&\! \ket{\Gamma_8^{(1)},7/2,-} \!&\! \ket{\Gamma_8^{(2)},5/2,-} \!&\! \ket{\Gamma_8^{(1)},7/2,+} \\
\hline
0.468 \!\!&\!\! 0.018\!-\!0.059 i & 0 & 0 & 0 & 0 & 0 & 0 \\
0.018\!+\!0.059 i \!\!&\!\! 0.026 & 0 & 0 & 0 & 0 & 0 & 0 \\
0      & 0      & 0.4678 \!\!&\!\! 0.018\!-\!0.059 i & 0 & 0 & 0 & 0 \\
0      & 0      & 0.018\!+\!0.059 i \!\!&\!\! 0.026 & 0 & 0 & 0 & 0 \\
0      & 0      & 0      & 0      & 0.468 \!\!&\!\! 0.018\!-\!0.059 i & 0 & 0 \\
0      & 0      & 0      & 0      & 0.018\!+\!0.059 i \!\!&\!\! 0.026 & 0 & 0 \\
0      & 0      & 0      & 0      & 0      & 0 & 0.468 \!\!&\!\! 0.018\!-\!0.059 i \\
0      & 0      & 0      & 0      & 0      & 0 & 0.018\!+\!0.059 i \!\!&\!\! 0.026 
\end{bmatrix}
\]

\[
\rho_{\Gamma_7} = 
\begin{bmatrix}
\ket{\Gamma_7,5/2,+} & \ket{\Gamma_7,7/2,-} & \ket{\Gamma_7,5/2,-} & \ket{\Gamma_7,7/2,+} \\
\hline
0.067 & 0.007\!-\!0.001i & 0 & 0 \\
0.007\!+\!0.001i & 0.035 & 0 & 0 \\
0      & 0      & 0.067 & 0.007\!-\!0.001i \\
0      & 0      & 0.007\!+\!0.001i & 0.035
\end{bmatrix}
\;\;\;
\rho_{\Gamma_6} =
\begin{bmatrix}
\ket{\Gamma_6,7/2,+} & \ket{\Gamma_6,7/2,-} \\
\hline
0.020 & 0 \\
0      & 0.020 
\end{bmatrix}
\]
Note that, because of the Shur lemma, $\rho$ has the same
block structure of the matrix $Z$.

We point out that the numbers reported in Table~I of the main text
correspond to the diagonal elements of the matrix $\rho$ in the basis
that diagonalizes $Z$ (that is not the same basis that diagonalizes $\rho$).

\subsection{G.~~~Many-body configuration probabilities UO$_2$}

\begin{figure}
	\begin{center}
		\includegraphics[width=8.7cm]{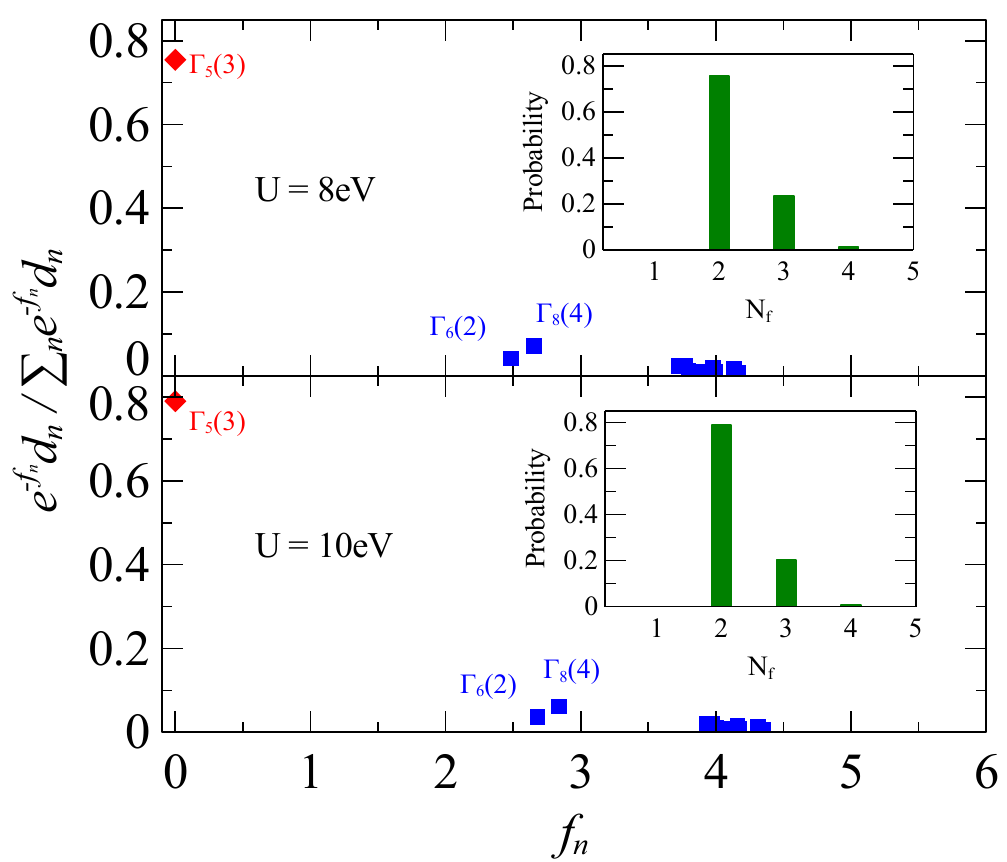}
		\caption{(Color online)
			Configuration probabilities of the eigenstates of the local reduced density
			matrix $\hat{\rho}_f\equiv e^{-\hat{F}}/\Tr[e^{-\hat{F}}]$
			of the $5f$ electrons shown as a function of the eigenvalues $f_n$ of $\hat{F}$
			at $V_{\text{eq}}\simeq 41\,\AA/\text{f.u.}$ for 2 different values of $U$.
			The labels of the irreducible representations and their respective degeneracies
			are expressed using the Koster notation.
			The corresponding occupation probabilities $\Tr\big[\hat{\rho}_f \hat{N}_f\big]$
			are shown in the insets.
		}
		\label{figure2}
	\end{center}
\end{figure}
In Fig.~\ref{figure2} are shown the eigenvalues
of the local reduced density matrix $\hat{\rho}_f$
of the U-$5f$ electrons --- which is formally obtained from
the full many-body density matrix of the system by tracing
out all of the degrees of freedom with the exception of the $5f$
local many-body configurations of the U atoms.
As in Ref.~\onlinecite{Our-PRX-supplemental}, $\hat{\rho}_f$ is represented as
$e^{-\hat{F}}/\Tr[e^{-\hat{F}}]$, and the corresponding eigenvalues
(configuration probabilities) are displayed
as a function of the corresponding eigenvalues $f_n$ of $\hat{F}$
(entanglement spectrum).
In the insets is shown also 
the histogram of occupation probabilities:
\be
P_N \equiv \Tr\big[\hat{\rho}_f \hat{N}_f\big]\,,
\ee
where $\hat{N}_f$ is the number operator of the U-$5f$ states.
The so obtained histogram is very similar for the 2 values of interaction strength $U$
considered.

Note that, because of the crystal field splittings,
the eigenstates of $\hat{\rho}_f$
generate irreducible representations
of the double $O$ point group of the U atom, 
whose transformation properties are represented in Fig.~\ref{figure2}
using the Koster notation.

Consistently with previous theoretical~\cite{RevModPhys-Actinides_oxides-supplemental,UO2-DFT+U-supplemental,UO2-Suzuki-Gamma5-supplemental}
and experimental~\cite{UO2-Amoretti-Gamma5-supplemental,UO2-CF-revisited-supplemental} studies,
we find that the most probable local configuration
is a $f^2$ $\Gamma_5$ triplet, which has probability $P_{\Gamma_5}^{f^2}\sim 0.8$
according to our calculations.
We point out that the $f^2$ many-body space contains $12$ $\Gamma_5$ representations.
Consequently, the above-mentioned most probable eigenspace
of $\hat{\rho}_f$, that we name $V_{\Gamma_5}^{f^2}$,
can not be determined exclusively by its symmetry properties, but has to be calculated.
For completeness, here we report the explicit representation of the states spanning
$V_{\Gamma_5}^{f^2}$ in the Fock basis generated by the single-particle states 
defined in Eq.~16 of the main text:
\bea
&&\{\ket{\alpha}\,|\,\alpha=1,..,14\}=\nonumber\\&&\;
\big\{
\ket{\Gamma_8^{(1)},5/2,+}, \ket{\Gamma_8^{(2)},7/2,-}, \ket{\Gamma_8^{(1)},5/2,-}, \ket{\Gamma_8^{(2)},7/2,+}, \ket{\Gamma_8^{(2)},5/2,+}, \ket{\Gamma_8^{(1)},7/2,-}, \ket{\Gamma_8^{(2)},5/2,-}, \ket{\Gamma_8^{(1)},7/2,+},\nonumber\\&&\;
\ket{\Gamma_7,5/2,+}, \ket{\Gamma_7,7/2,-}, \ket{\Gamma_7,5/2,-}, \ket{\Gamma_7,7/2,+},
\ket{\Gamma_6,7/2,+}, \ket{\Gamma_6,7/2,-}
\big\}\,.
\eea
Within the following notation:
\be
\ket{\Gamma_5,n}=\sum_{\alpha=2}^{14}\sum_{\beta=1}^{\alpha-1}
M_{\alpha\beta}^n \,\cc_\alpha\cc_\beta\,\ket{0}
\;,\qquad\quad n=1,2,3\,,
\ee
the states $\ket{\Gamma_5,n}$ are specified by the following coefficients:
\bea
M^1_{21}&=&-0.004+ 0.042 i
\nonumber\\
M^1_{3\beta}&=&(0, 0)
\nonumber\\
M^1_{4\beta}&=&(0, 0, 0.005-0.01 i)
\nonumber\\
M^1_{5\beta}&=&(-0.154+ 0.25 i, 0.03-0.049 i, -0.074+0.808 i,  0.008-0.085i)
\nonumber\\
M^1_{6\beta}&=&(0.012-0.019i, -0.004+0.006i, 0.012-0.134i, -0.002+0.021i, -0.005+0.01i)
\nonumber\\
M^1_{7\beta}&=&(0.097-0.195j, -0.01+0.021i, -0.154+0.25i, 0.03-0.049i, 0, 0)
\nonumber\\
M^1_{8\beta}&=&(-0.016+0.032i, 0.003-0.005i, 0.012-0.019i,-0.004+0.006i, 0, 0, 0.004-0.042i)
\nonumber\\
M^1_{9\beta}&=&(-0.016+0.171i, 0.007-0.073i, 0, 0, 0.038-0.061i, -0.016+0.026i, 0.012-0.024i,
\nonumber\\&&-0.005+0.01i)
\nonumber\\
M^1_{10\beta}&=&(0.005-0.06i, -0.002+ 0.022i, 0, 0, -0.013+ 0.021i, 0.005-0.008i, -0.004+ 0.008i,
\nonumber\\&&  0.002-0.003i, 0)
\nonumber\\
M^1_{11\beta}&=&(0, 0, 0.021-0.041i, -0.009+ 0.018i, -0.009+ 0.099i, 0.004-0.042i, 0.038-0.061i,
\nonumber\\&&  -0.016+ 0.026i, 0,0)
\nonumber\\
M^1_{12\beta}&=&(0, 0,-0.007+ 0.014i, 0.003-0.005i, 0.003-0.035i, -0.001+ 0.013i, -0.013+ 0.021i,
\nonumber\\&&  0.005-0.008i, 0, 0, 0)
\nonumber\\
M^1_{13\beta}&=&(0.004-0.008i, 0.001-0.003i, 0.013-0.021i, 0.004-0.006i, 0, 0, 0.005-0.06i, 0.002-0.018i,
\nonumber\\&&-0.007+ 0.014i, 0.004-0.007i, 0.011-0.017i, -0.006+ 0.009i)
\nonumber\\
M^1_{14\beta}&=&(-0.013+ 0.021i,-0.004+ 0.006i, 0.003-0.034i,  0.001-0.01i, 0.007-0.014i, 0.002-0.004i,
\nonumber\\&&  0, 0, -0.011+ 0.017i,  0.006-0.009i, -0.005+ 0.056i, 0.003-0.030i,0)
\eea

\bea
M^2_{21}&=& -0.015-0.007j  
\nonumber\\
M^2_{3\beta}&=& (0, 0)
\nonumber\\
M^2_{4\beta}&=&  (0, 0, -0.030-0.034i) 
\nonumber\\
M^2_{5\beta}&=& (0.069+ 0.083i, -0.013-0.016i, -0.279-0.14i,  0.029+ 0.015i)
\nonumber\\
M^2_{6\beta}&=& (-0.005-0.006i, 0.002+ 0.002i,  0.046+ 0.023i, -0.007-0.004i, 0.03+0.034i)  
\nonumber\\
M^2_{7\beta}&=& (-0.574-0.655i, 0.061+ 0.069i, 0.069+ 0.083i, -0.013-0.016i, 0, 0)
\nonumber\\
M^2_{8\beta}&=&  ( 0.095+ 0.109i, -0.015-0.017i, -0.005-0.006i, 0.002+ 0.002i, 0, 0, 0.015+ 0.007i) 
\nonumber\\
M^2_{9\beta}&=& (-0.059-0.03i, 0.025+ 0.013i, 0, 0, -0.017-0.02i, 0.007+ 0.009i, -0.07-0.08i,  0.03+ 0.034i)
\nonumber\\
M^2_{10\beta}&=&( 0.021+ 0.01i, -0.008-0.004i, 0, 0, 0.006+ 0.007i, -0.002-0.003i, 0.025+ 0.028i, -0.009-0.01i, 0)
\nonumber\\
M^2_{11\beta}&=& (0, 0, -0.122-0.139i, 0.052+ 0.059i, -0.034-0.017i, 0.014+ 0.007i, -0.017-0.02i,  0.007+ 0.009i, 0, 0)
\nonumber\\
M^2_{12\beta}&=& (0, 0, 0.043+ 0.048i, -0.016-0.018i, 0.012+ 0.006i, -0.004-0.002i, 0.006+ 0.007i,  -0.002-0.003i, 0, 0, 0)
\nonumber\\
M^2_{13\beta}&=& (-0.024-0.028i, -0.007-0.008i, -0.006-0.007i, -0.002-0.002i, 0, 0, 0.021+ 0.01i, 0.006+ 0.003i,
\nonumber\\&&  0.04+ 0.046i, -0.021-0.024i, -0.005-0.006i, 0.003+ 0.003i)
\nonumber\\
M^2_{14\beta}&=& ( 0.006+ 0.007i, 0.002+ 0.002i,  0.012+ 0.006i, 0.004+ 0.002i, -0.042-0.048i, -0.013-0.015i, 0, 0,
\nonumber\\&& 0.005+ 0.006i, -0.003-0.003i, -0.019-0.01i, 0.01+ 0.005i, 0)
\eea

\bea
M^3_{21}&=& 0.018+ 0.001i  
\nonumber\\
M^3_{3\beta}&=& ( 0, 0 )
\nonumber\\
M^3_{4\beta}&=& ( 0, 0, -0.013-0.006i )  
\nonumber\\
M^3_{5\beta}&=& (-0.513-0.280i,  0.101+ 0.055i, 0.35+ 0.011i, -0.037-0.001i )  
\nonumber\\
M^3_{6\beta}&=& ( 0.039+ 0.021i, -0.013-0.007i, -0.058-0.002i, 0.009+ 0.000i, 0.013+ 0.006i )  
\nonumber\\
M^3_{7\beta}&=& (-0.244-0.117i, 0.026+ 0.012i, -0.513-0.280i, 0.101+ 0.055i, 0, 0 )  
\nonumber\\
M^3_{8\beta}&=& ( 0.04+ 0.019i, -0.006-0.003i, 0.039+ 0.021i, -0.013-0.007i, 0, 0, -0.018-0.001i )  
\nonumber\\
M^3_{9\beta}&=& ( 0.074+ 0.002i,  -0.031-0.001i, 0, 0, 0.126+ 0.069i, -0.053-0.029i, -0.03-0.014i, 0.013+ 0.006i)
\nonumber\\
M^3_{10\beta}&=& (-0.026-0.001i, 0.01, 0, 0, -0.044-0.024i, 0.016+ 0.009i, 0.01+ 0.005i, -0.004-0.002i, 0)
\nonumber\\
M^3_{11\beta}&=& (0, 0, -0.052-0.025i, 0.022+ 0.010i, 0.043+ 0.001i, -0.018-0.001i, 0.126+ 0.069i, -0.053-0.029i, 0,0)
\nonumber\\
M^3_{12\beta}&=& (0, 0, 0.018+ 0.009i, -0.007-0.003i, -0.015, 0.006, -0.044-0.024i, 0.016+ 0.009i, 0, 0, 0)
\nonumber\\
M^3_{13\beta}&=& (-0.010-0.005i, -0.003-0.001i, 0.044+ 0.024i, 0.013+ 0.007i, 0, 0, -0.026-0.001i, -0.008,
\nonumber\\&& 0.017+ 0.008i, -0.009-0.004i, 0.036+ 0.02i, -0.019-0.01i)  
\nonumber\\
M^3_{14\beta}&=& (-0.044-0.024i, -0.013-0.007i, -0.015, -0.004, -0.018-0.009i, -0.005-0.003i, 0, 0,
\nonumber\\&& -0.036-0.02i, 0.019+ 0.01i, 0.024+ 0.001i, -0.013,0)\,.
\eea

We observe that the remaining probability weight, which is not negligible, 
is distributed mostly among $f^3$ configurations,
whose Koster symbols are displayed explicitly in Fig.~\ref{figure2}
for the most probable multiplets.

\end{document}